# Vapor-Liquid-Solid Growth of Monolayer MoS$_2$ Nanoribbons

*Shisheng Li[1,2], Yung-Chang Lin[3], Wen Zhao[4], Jing Wu[5], Zhuo Wang[1,2], Zehua Hu[1,2], Youde Shen[6], Dai-Ming Tang[7], Junyong Wang[1,2], Qi Zhang[2], Hai Zhu[8], Leiqiang Chu[1,2], Weijie Zhao[1,2], Chang Liu[9], Zhipei Sun[10], Takaaki Taniguchi[7], Minoru Osada[7], Wei Chen[1,2,8], Qing-Hua Xu[8], Andrew Thye Shen Wee[1,2], Kazu Suenaga[3] Feng Ding[4,11] and Goki Eda[1,2,8]*

[1] Centre for Advanced 2D Materials, National University of Singapore, 117542, Singapore

[2] Department of Physics, National University of Singapore, 117542, Singapore

[3] Nanotube Research Center, National Institute of Advanced Industrial Science and Technology, Tsukuba 305-8564, Japan

[4] Center for Multidimensional Carbon Materials, Institute for Basic Science, Ulsan, 689-798, Republic of Korea

[5] Institute of Materials Research and Engineering, Agency for Science, Technology, and Research, 138634, Singapore

[6] Department of Electrical and Computer Engineering, National University of Singapore, 117583, Singapore

[7] International Center for Materials Nanoarchitectonics, National Institute for Materials Science, Tsukuba 305-0044, Japan

[8] Department of Chemistry, National University of Singapore, 117543, Singapore

[9] Institute of Metal Research, Chinese Academy of Sciences, Shenyang, 110016, China

[10] Department of Micro and Nanosciences, Aalto University, Espoo FI-02150, Finland

[11] School of Materials Science and Engineering, Ulsan National Institute of Science and Technonlogy, Ulsan, 689-798, Republic of Korea

Correspondence and requests for materials should be addressed to G.E. (email: g.eda@nus.edu.sg), S.L. (email: shishengli1108@gmail.com)

**Chemical vapor deposition (CVD) of two-dimensional (2D) materials such as monolayer MoS$_2$ typically involves conversion of vapor-phase precursors to a solid product in a process that may be described as a vapor-solid-solid (VSS) mode. Here, we report the first demonstration of vapor-liquid-solid (VLS) growth of monolayer MoS$_2$ yielding highly crystalline ribbon-shaped structures with a width of a few tens of nanometers to a few micrometers. The VLS growth mode is triggered by the reaction between molybdenum oxide and sodium chloride, which results in the formation of molten Na-Mo-O droplets. These droplets mediate the growth of MoS$_2$ ribbons in the "crawling mode" when saturated with sulfur on a crystalline substrate. Our growth yields straight and kinked ribbons with a locally well-defined orientation, reflecting regular horizontal motion of the liquid droplets during growth. Using atomic-resolution scanning transmission electron microscopy (STEM) and second harmonic generation (SHG) microscopy, we show that the ribbons are homoepitaxially on monolayer MoS$_2$ surface with predominantly 2H- or 3R-type stacking. These findings pave the way to novel devices with structures of mixed dimensionalities.**

Recent advances in the chemical vapor deposition (CVD) of two-dimensional (2D) transition metal dichalcogenides (TMDs) on various substrate surfaces[1,2] have opened up new prospects for the exploration of their fundamental physical properties[3] and practical device implementation schemes[4]. However, the range of structures that can be controllably synthesized by the current methods in terms of morphology, spatial selectivity, crystal orientation, layer number, and chemical composition is still limited. Development of versatile growth methods is essential to enabling the realization of highly integrated electronic and photonic devices based on these materials.

While various CVD-based growth techniques for TMDs have been demonstrated to date[1,2], they generally involve similar microscopic mechanism where the gas/vapor phase precursors are converted to a solid-state product via surface adsorption, surface diffusion, and bond formation[5]. This growth mode, which may be referred to as vapor-solid-solid (VSS) type, is common in the CVD growth of not only 2D TMDs[6-9] but also other ultrathin chalcogenides[10] as well as graphene[11-13]. Since the precursors are uniformly supplied to the nuclei in the 2D growth plane, this type of growth yields structures with a characteristic crystal shape determined primarily by the inherent free energy of the crystal edges and surface diffusion kinetics[14].

An alternative growth mode known as the vapor-liquid-solid (VLS) growth is an attractive approach to introducing lateral control in 2D crystal growth and achieving direct bottom-up synthesis of integration-ready nanostructures. In VLS mode, one-dimensional (1D) nanostructures are produced by precipitation from supersaturated catalytic liquid droplets[15-17]. VLS growth has been observed for various van der Waals layered compounds including BN[18], NiCl$_2$[19], SnS$_2$[20] and Bi$_2$Se$_3$[21], but the growth products are often randomly oriented multilayered nanotubes or nanoribbons, typically consisting of tens to hundreds of monolayers. Meanwhile, recent demonstrations of graphene[22] and 2D oxide[23] growth on liquid metal surfaces, and monolayer MoSe$_2$ growth on molten glass[24] highlight the unique potential of liquid-mediated synthesis techniques. Suzuki et al.[25] succeeded in VLS-like growth of laterally oriented graphene nanoribbons from molten nickel. Their observation suggests that non-tubular, atomically thin structures can be synthesized in VLS mode and offers prospects towards the controlled 1D growth of layered compound semiconductors such as TMDs in their monolayer limit.

In this article, we report in-plane VLS growth of monolayer $MoS_2$ ribbons with an average width of hundreds of nanometers on crystalline surfaces. We discover that alkali metal halide reacts with transition metal oxide precursors to form molten droplets that crawl on the substrate surface and mediate the highly anisotropic growth. We demonstrate that the VLS mode allows homoepitaxial growth of ribbons on a pre-grown monolayer $MoS_2$ support layer, yielding unique 1D-on-2D structures. We also show that the alignment of the ribbons is largely determined by the orientation of the underlying crystal, similar to the guided horizontal growth of nanotubes[26] and nanowires[27]. Our results provide insight into the distinctly new growth mode of 2D $MoS_2$ and offer prospects for their nanoelectronic device implementations.

**VLS growth of $MoS_2$ nano- and micro-ribbons on a NaCl crystal**

We conducted salt-assisted CVD[9] of monolayer $MoS_2$ on a freshly cleaved surface of a NaCl single crystal using powder $MoO_3$ and S as precursors (See Methods and Supplementary Fig. 1 for details). Figure 1a and b show optical and atomic force microscope (AFM) images of monolayer $MoS_2$ grown on the NaCl surface (see also Supplementary Fig. 2 for more images). In stark contrast with the typical triangular crystals, straight and kinked narrow ribbons with a width of a few tens of nanometers to a few micrometers, and a length ranging from a few to tens of micrometers are grown (see Supplementary Fig. 2i for width distribution). The step height of a majority of these ribbons is ~ 0.8 nm, which is consistent with the thickness of monolayer $MoS_2$[28] (Supplementary Fig. 3). The optical images show that most ribbons are terminated with a particle as indicated by the circles in Figure 1a-c. We note that the size of these particles consistently matches the width of the ribbons. These are characteristic features of nanostructures resulting from the "crawling mode"[26,27,29] of VLS growth where liquid droplets crawl on the surface as the reaction product is precipitated laterally. These ribbons are locally aligned but exhibit occasional regular kinks, suggesting that the growth is guided either by the substrate or the crystal facets of $MoS_2$ (Figure 1 a and d, Supplementary Fig. 2).

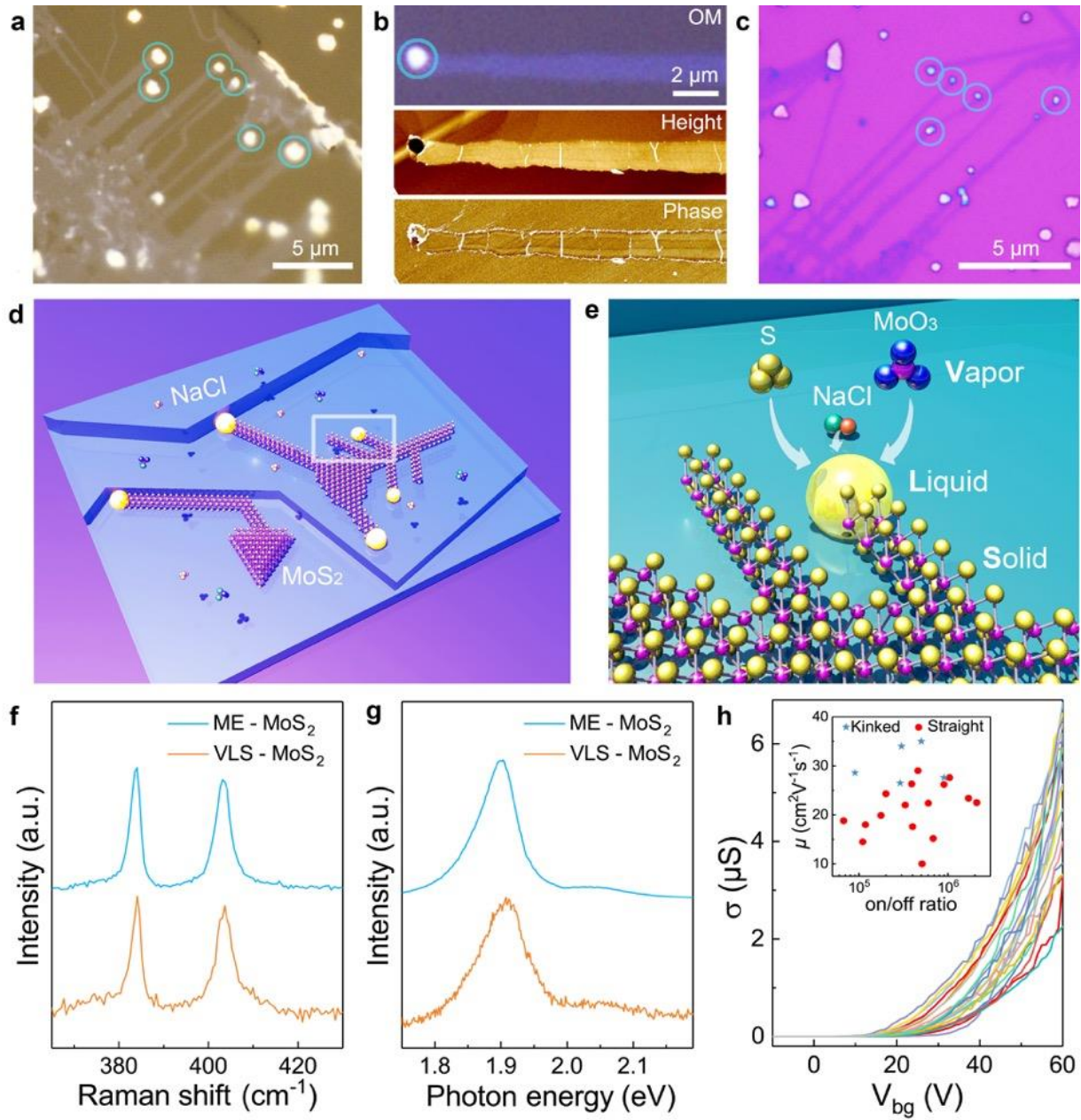

**Figure 1 | Morphology and properties of VLS-MoS$_2$ ribbons grown on a NaCl single crystal.** (a) Optical image of MoS$_2$ ribbons grown on a NaCl crystal. (b) Optical and corresponding AFM images of MoS$_2$ ribbons. (c) Optical image of MoS$_2$ ribbons transferred on a SiO$_2$/Si substrate. Circles highlight the terminal particles attached to the MoS$_2$ ribbons. (d, e) Schematic illustrations of VLS growth of MoS$_2$ ribbons on a NaCl crystal surface. The reaction involves: 1) formation of liquid phase Na-Mo-O in small droplets; 2) dissolution of sulfur in the droplet; 3) horizontal growth of MoS$_2$ ribbons and lateral displacement of the droplet. (f) Raman and (g) PL spectra of as-transferred MoS$_2$ ribbons and mechanically exfoliated (ME) monolayer MoS$_2$ on SiO$_2$/Si substrates. (h) Transport properties of MoS$_2$ nanoribbon FETs. Inset shows the plot of mobility versus current on/off ratio for all measured MoS$_2$ nanoribbon FETs.

To understand the possible reaction routes involving liquid-phase intermediate compound, we conducted thermogravimetric analysis (TGA) of the growth precursors, X-ray diffraction (XRD) and energy dispersive X-ray (EDX) analysis on the reaction products. While our growth temperature (~ 700 °C) is below the typical melting point of $MoO_3$ (795 °C) and NaCl (801 °C) in atmospheric pressure, these compounds gradually sublime at the growth temperature as evidenced by their deposition on $SiO_2$/Si substrate in a control experiment (Supplementary Fig. 4). TGA of $MoO_3$-NaCl mixture (Supplementary Fig. 5) shows a prominent weight loss at ~ 550 °C in contrast to pure $MoO_3$, which shows weight loss only above 800 °C. These results indicate that NaCl chemically reacts with $MoO_3$ at temperatures well below the melting point of the two compounds. Further, XRD study of the reaction product reveals that the resulting compound contains $Na_2Mo_2O_7$, which has a low melting point of 605 °C[30] and is a liquid at the growth temperature (Supplementary Figs. 4 and 5). A possible reaction route[31] is

$$3\ MoO_3\ (s) + 2\ NaCl\ (s) \rightarrow Na_2Mo_2O_7\ (l) + MoO_2Cl_2\ (g). \qquad (1)$$

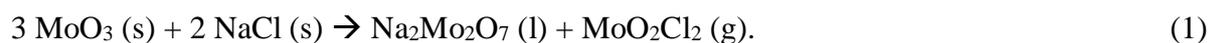

We experimentally found that the liquid phase $Na_2Mo_2O_7$ indeed yields $MoS_2$ when exposed to sulfur vapor at the growth temperature (Supplementary Fig. 5). From EDX analysis, the particles terminating the ribbons are found to contain molybdenum, sodium, oxygen, sulfur, and a small amount of chlorine (Supplementary Fig. 6). These results further support the validity of the above reaction route and VLS growth as schematically illustrated in Figures 1d and e.

Sodium plays a key role in lowering the melting point of $MoO_3$ precursor and enabling the VLS mode. This is similar to the role of gold in the VLS growth of Si and III-V semiconductor nanowires where gold forms a eutectic with the precursor element and facilitate liquid-mediated growth at temperatures lower than required in other growth modes[15,17,27,29,32]. Alloys of alkali metal, transition metal and oxygen have similar melting points and can effectively mediate VLS growth of TMDs[33]. We found that reacting pre-deposited sodium molybdate ($Na_2MoO_4$), which is also a liquid at 700 °C, with sulfur vapor yields monolayer $MoS_2$ ribbons, triangular crystals, and continuous thin films (Supplementary Figs. 7 and 8). This observation reveals a few important aspects of the reaction. First, chlorine is not an essential element in triggering the VLS mode. Second, the growth can be achieved in absence of $MoO_3$ vapor. Third, the liquid phase precursor can also mediate 2D growth yielding a continuous film. Further, we tested the general applicability of this growth method by conducting the growth of other monolayer TMDs ($MoSe_2$, $WS_2$, $WSe_2$, $MoTe_2$, $WTe_2$) and their heterostructure ($MoS_2$/$MoSe_2$) by reacting $Na_2MoO_4$ and $Na_2WO_4$ with chalcogen vapors above their melting point (Supplementary Figs. 8 and 9). In all cases, the desired materials could be grown, indicating the versatility of this approach.

Raman spectrum of $MoS_2$ ribbons transferred on a $SiO_2$/Si substrate shows characteristic $E_{2g}^1$ and $A_{1g}$ peaks at 384.2 and 403.7 $cm^{-1}$ with a peak separation of 19.5 $cm^{-1}$ (Figure 1f). The ribbons also exhibited characteristic excitonic photoluminescence (PL) peaks at 1.9 and 2.05 eV (Figure 1g). These features are nearly identical to those of mechanically exfoliated (ME-) monolayer $MoS_2$, indicating that our VLS growth does not introduce substantial defects, strain and doping beyond the level of exfoliated materials. We further verified this by investigating the electronic quality of the nanoribbons by evaluating their field-effect mobility. We tested 21 two-terminal field-effect transistors (FET) based on straight and kinked $MoS_2$ nanoribbons and found that they exhibit typical n-type transfer characteristics with on/off ratio of ~$10^5$ and electron field-effect mobility range of 10-35 $cm^2V^{-1}s^{-1}$, which is comparable to the performance of monolayer $MoS_2$ prepared by mechanical exfoliation[34] and metal-organic CVD[2]. Both types of ribbons exhibited similar average performance, indicating that kinks do not represent major structural defects detrimental to carrier transport at room temperature (Figure 1h).

**VLS epitaxy of $MoS_2$ nano- and micro-ribbons on monolayer $MoS_2$**

In order to test the feasibility of VLS growth on other surfaces, we further conducted salt-assisted growth of $MoS_2$ ribbons on a continuous film of monolayer $MoS_2$ pre-grown on a $SiO_2$/Si substrate (See Supplementary Information for detailed method). Figure 2a and Supplementary Fig. 10 show the unique morphology of kinked ribbons on monolayer $MoS_2$ support layer. Majority of these ribbons were found to be monolayers (Supplementary Fig. 11a-c). The orientation of the ribbons is locally well defined and their predominant growth directions are separated by ~ 60 ° or ~ 120 ° (Figure 2a). Most ribbons are terminated by a particle similar to the case of growth on NaCl single crystal (Figure 2a and b).

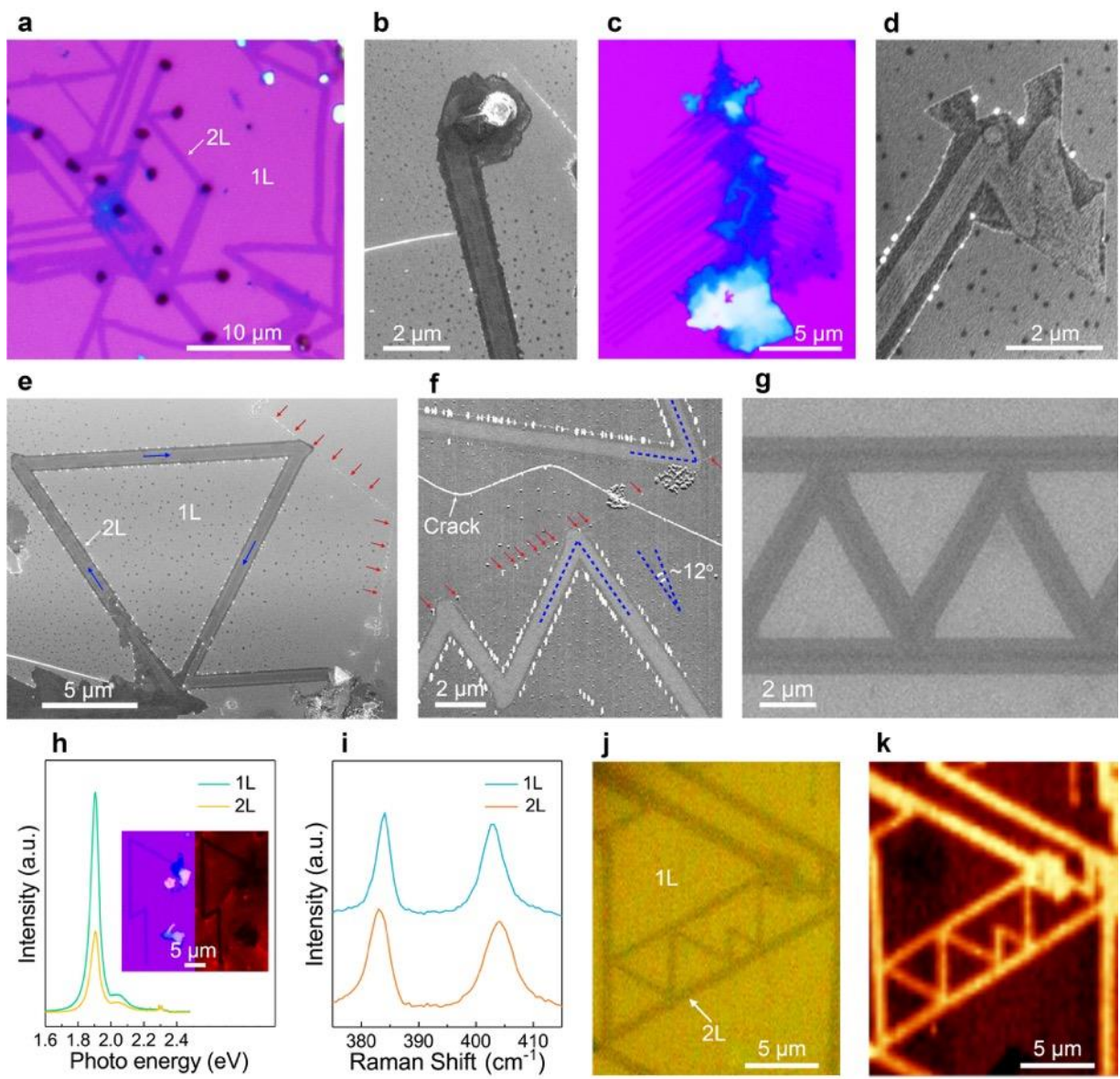

**Figure 2 | VLS homoepitaxy of MoS$_2$ ribbons on monolayer MoS$_2$.** (a) Optical image of ribbons on a monolayer MoS$_2$ film. (b) SEM image of a particle attached to the ribbon terminal. (c) Parallel arrays of ribbons extending from a MoS$_2$ crystal island. (d) SEM image of the terminal morphology of a ribbon in the array. (e) SEM and (f) AFM images of ribbons showing kinks at the line features (indicated by the arrows that point to the aggregated particles) of the underlying MoS$_2$. These line features are most likely the grain boundary of the underlying monolayer MoS$_2$ (See Supplementary Information for further discussions). (g) Zigzag-shaped MoS$_2$ ribbon which resulted from a liquid droplet trapped in between two parallel ribbons. (h) Micro-PL spectra collected from the underlying monolayer and an overgrown ribbon. Inset are optical and fluorescence images of the structure. (i) Raman spectra of underlying monolayer and an overgrown ribbon. (j) Optical image and (k) corresponding Raman A$_{1g}$ intensity map of MoS$_2$ ribbons grown on monolayer MoS$_2$.

Figure 2c shows an optical image of arrays of numerous parallel nanoribbons extending from an island of multilayered $MoS_2$. This morphology suggests that the alignment of the ribbons is induced by the orientation of the underlying $MoS_2$ crystal. Figure 2d shows another frequently observed ribbon termination where a faceted 2D flake is formed instead of a particle. Absence of a particle suggests that the precursor was fully consumed during growth. Careful examination of the ribbon morphologies by scanning electron microscope (SEM) (Figure 2e) and AFM (Figure 2f) reveals that the kinks often appear along straight line features of the underlying monolayer $MoS_2$. Particles tend to aggregate along these lines as indicated by the arrows in Figure 2e and f. Based on a series of measurements, we found that these lines correspond to the grain boundary (GB) of the underlying $MoS_2$ (See Supplementary Fig. 11d-f for further discussions). The features of the ribbons around the kinks reveal the history of the liquid droplet migration. For example, the morphology of the ribbon in Figure 2e suggests that the droplet initially moved towards the upper left region before encountering a surface perturbation and changing its direction by ~ 120 º as marked by the arrows. The growth of this ribbon was terminated after the droplet took two additional turns, one at a GB and another at a ribbon edge. It is also worth highlighting that each kink shows a round corner, which is indicative of the liquid state of the precursor at the growth temperature (Supplementary Figs. 6 and 11b). Figure 2f shows an AFM phase image of two ribbons grown on two adjacent grains of the underlying $MoS_2$ film. In both grains, the ribbon takes a ~120 º turn at the GB but the orientations of the ribbons in the two adjacent grains are offset by ~12 º. The difference in the ribbon orientation is most likely caused by the orientation difference of the underlying $MoS_2$ grains. Another commonly observed scenario involves a droplet trapped in between two parallel $MoS_2$ ribbons as shown in Figure 2g. In this case, the liquid droplet "bounces" back and forth with ~ 120 º turns between the parallel ribbons, resulting in a peculiar zigzag-shaped ribbon.

Micro-PL spectroscopy and fluorescence imaging show that the characteristic excitonic direct gap emission of monolayer $MoS_2$ is significantly quenched along the ribbons (Figure 2h). Quenching of PL suggests that there is a strong interaction between the upper and lower $MoS_2$ layers, resulting in indirect band gap formation in the bilayer regions[35,36]. Raman spectra show that the $E_{2g}^1$ and $A_{1g}$ peaks of the bilayer regions exhibit characteristic softening and stiffening as observed in commensurately stacked bilayers, further indicating strong interlayer interaction (Figure 2i). Figure 2j and k are optical image and corresponding integrated intensity map of Raman $A_{1g}$ peak of the nanoribbons, respectively, showing distinctly stronger and uniform intensity in the bilayer region, highlighting the structural uniformity of the epitaxial layer.

**Stacking order of 1D-on-2D MoS$_2$ structures**

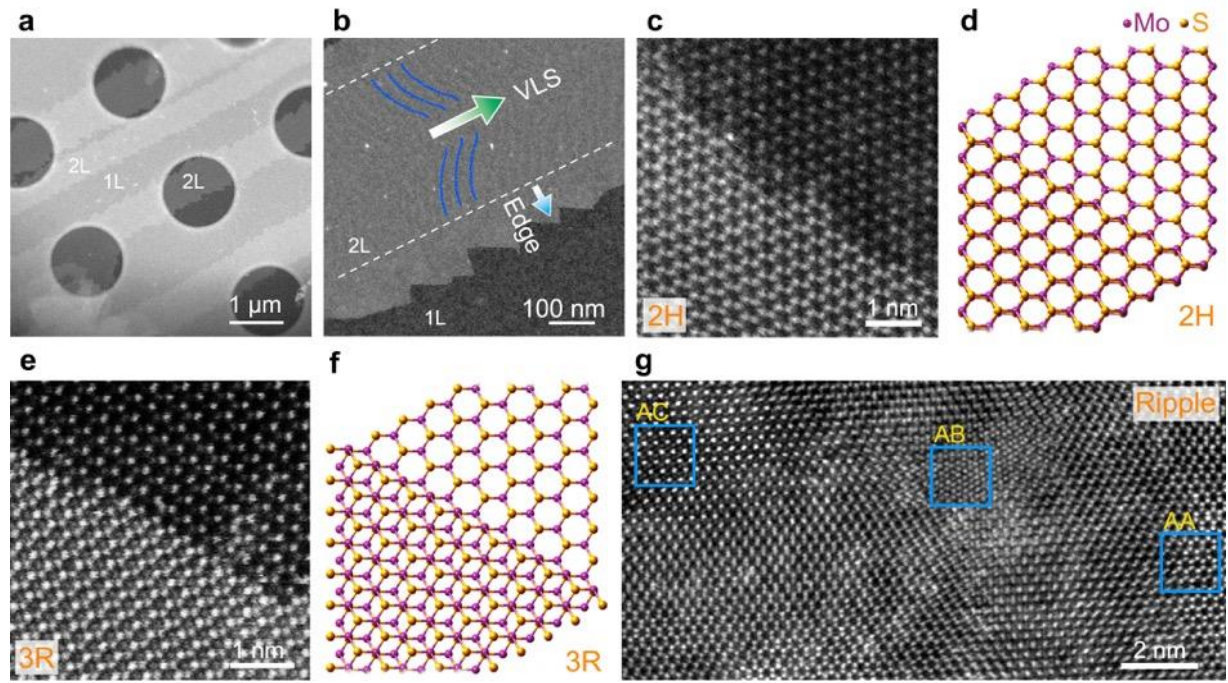

**Figure 3 | STEM images of homoepitaxially grown MoS$_2$ ribbons on monolayer MoS$_2$.** (a) ADF-STEM image of four parallel MoS$_2$ ribbons. The ribbons appear in bright contrast. (b) High-magnification TEM image showing the zigzag-shaped edges and fishbone-like contrast at the center of the ribbon. The arrows indicate edge regions where the contrast is uniform. (c) STEM image and (d) corresponding atomic configuration of the edge of a ribbon showing the boundary between 2H-stacked bilayer and supporting monolayer. (e) STEM image and (f) corresponding atomic configuration of an edge of a ribbon showing the boundary between 3R-type bilayer and supporting monolayer. Purple balls in (d) and (f) represent Mo atoms, and yellow balls represent S atoms. (g) STEM image showing spatially varying stacking sequence in the center region of a ribbon. AA, AB, and AC represent three different local stacking sequences.

To further investigate the stacking order of the epitaxially grown ribbons on monolayer MoS$_2$, we conducted scanning transmission electron microscopy (STEM) imaging of the samples. Figure 3a and b show a low-magnification annular dark field (ADF) STEM image of ribbons on a monolayer $MoS_2$ support, revealing their rough zigzag-shaped edges. Atomic resolution STEM images near the edges of the bilayer regions reveals that two types of stacking orders, namely 2H and 3R, are predominant. Figure 3c and 3d show STEM image and corresponding atomic model of bilayer $2H-MoS_2$. 2H-type bilayer is characterized by a stacking sequence where the two layers have 180 º relative orientation, restoring inversion symmetry. The metal (chalcogen) atoms in the upper layer are aligned with the chalcogen (metal) atoms in the lower layer, thereby forming a hollow hexagonal structure when viewed from the c-axis direction (Figure 3d). Figure 3e and 3f show STEM image and corresponding atomic model of the 3R-type bilayer. This 3R stacking is characterized by a stacking sequence where the two layers have the same orientation but the top layer is shifted by $1/\sqrt{3}$ lattice constant along the <110> direction. These structures can be readily identified from the ADF contrast. The observed predominance of commensurate stacking is a direct evidence of epitaxial growth.

Detailed analysis of the atomic structure across the width of the ribbons reveals that the stacking order is spatially varied. Unlike the edge regions, which are uniform in contrast (blue arrow in Figure 3b), the center region of the ribbons exhibits peculiar fishbone-like periodic fringes (highlighted by blue lines in Figure 3b). These fringes suggest undulations of the layers with local strain and spatially varying interlayer interactions[37]. Figure 3g shows the atomic-resolution STEM image of the center of the ribbon where the dominant stacking is of 2H-type. It can be seen that the stacking order changes across the ribbon from AC (left square), AB (middle square) to AA (right square) over several nanometers. Careful examination of the signal intensity reveals that the relative orientation of the two layers remains the same while the layers are locally displaced with respect to one another (Supplementary Fig. 12). Such an evolution is indicative of the presence of strain and local separation of the layers[37]. As discussed below, the ripples may be induced a compressive force exerted by the liquid droplet during growth. It is worth noting that we did not find any evidence of Na and Cl impurities in the crystal lattice of $MoS_2$ based on the atomically resolved Z-contrast imaging of our samples.

**Probing stacking order by non-linear optics**

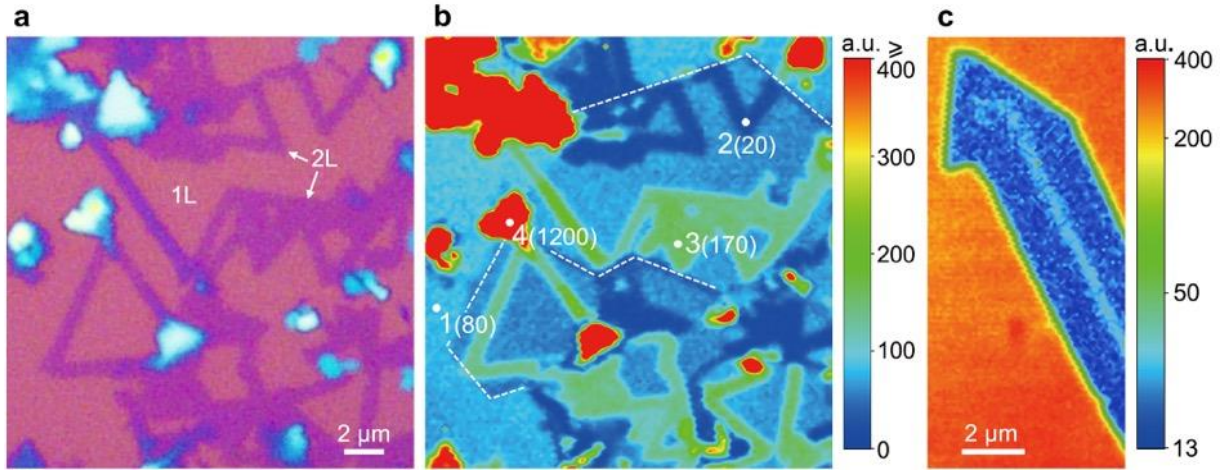

**Figure 4 | SHG microscopy of homoepitaxially grown MoS₂ ribbons on monolayer MoS₂.** (a) Optical image and (b) corresponding SHG image of MoS$_2$ ribbons and thick layers grown on monolayer MoS$_2$. The dark blue regions (point 2) in (b) represent ribbons exhibiting SHG signals weaker than the underlying monolayer (point 1). On the other hand, green regions (point 3) correspond to ribbons with SHG signals stronger than the underlying monolayer. The GBs of the underlying MoS$_2$ layer are identified from the contrast of SHG signal and highlighted with a dashed line. (c) SHG image of a ribbon with non-uniform SHG signal. The overall signal is weaker than the background suggesting that it is a bilayer with predominantly 2H phase. The signal is slightly enhanced in the center region of the ribbon due to the spatially varying of stacking sequence and weaker interlayer interaction.

Stacking order has a significant impact on the non-linear optical properties of MoS$_2$[38-41]. In order to further investigate the spatial variations of stacking sequence in the homoepitaxially grown ribbons, we conducted second harmonic generation (SHG) micro-spectroscopy of our samples. It is known that 2H-MoS$_2$ bilayers exhibit low second order non-linear optical susceptibility due to inversion symmetry and their SHG signals are typically two orders of magnitude smaller than that of non-centrosymmetric monolayer MoS$_2$[40,41]. In contrast, 3R-MoS$_2$ lacks inversion symmetry and exhibits intense SHG signals even in multilayer form[42]. Figure 4a and b show optical image and corresponding SHG intensity map of the homoepitaxially grown ribbons on monolayer MoS$_2$. While all the bilayer regions are uniform in contrast in the bright-field image, it is evident that different bilayer ribbons exhibit distinct non-linear optical response. Taking the SHG signals from the support monolayer as a reference (point 1 in Figure 4b), it can be seen that some ribbons exhibit (point 2) one order of magnitude weaker SHG signals while others show nearly twice enhanced signals (point 3) (Supplementary Fig. 13). This quenching and enhancement of SHG signals can be attributed to 2H and 3R-type stacking of the ribbons, respectively, further verifying the epitaxial nature of the VLS growth. Figure 4c shows the SHG intensity map of a wide ribbon with predominantly 2H stacking based on average SHG intensity. The center region of the ribbon exhibits slightly enhanced signals, indicating weaker interlayer interactions. This observation is consistent with the STEM analysis that the symmetry and interlayer interaction vary over nanometer length scales in the center region of the ribbons due to strain and nanoripples (Figure 3g).

**Mechanism of VLS growth of MoS$_2$ ribbons**

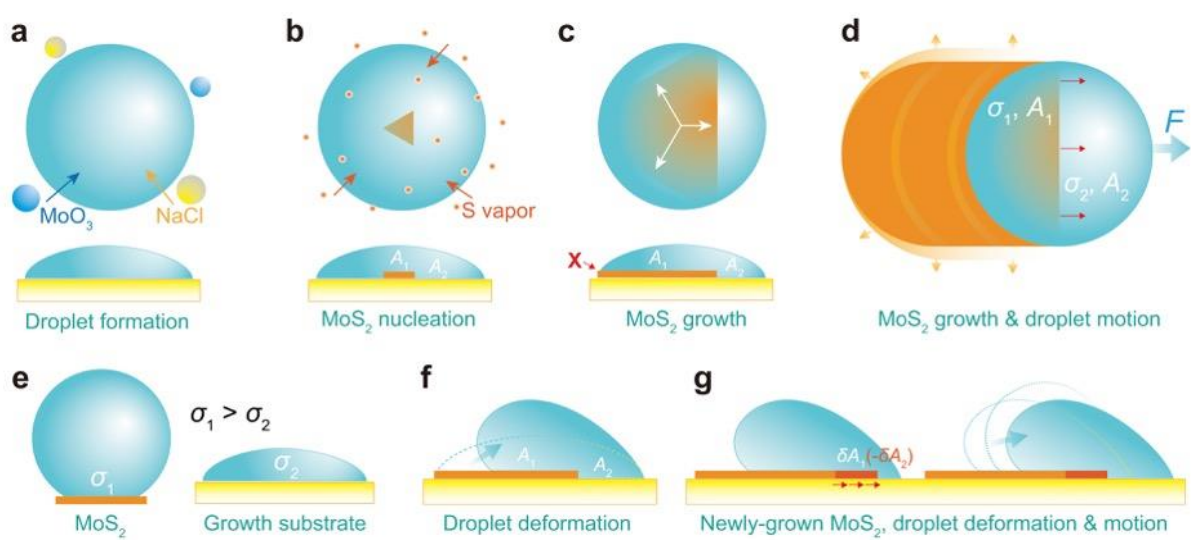

**Figure 5 | Schematic illustration of the mechanism of ribbon formation.** The VLS growth involves: (a) liquid droplet formation due to vapor phase reaction of MoO$_3$ and NaCl; (b) nucleation of MoS$_2$ flake at the droplet-substrate interface; (c) lateral growth of MoS$_2$ flake; (d) lateral horizontal displacement of the droplet and continuous growth of MoS$_2$ ribbon. (e) The droplet motion is induced by the interface energy difference between the droplet-MoS$_2$ ribbon ($\sigma_1$) and droplet-substrate ($\sigma_2$) interfaces. The droplet exhibits poorer wettability on MoS$_2$ ribbon than on growth substrate for $\sigma_1 > \sigma_2$. (f) De-wetting of droplet from "Position X" of Figure c to minimize the total interface free energy ($G_{\text{interface}}$), a quasi-equilibrium state is achieved. (g) Newly-grown MoS$_2$ ribbon leads to the increase of $G_{\text{interface}}$. The droplet rolls off the nanoribbon to minimize the $G_{\text{interface}}$ and restore the quasi-equilibrium state as illustrated in Figure f.

What drives the horizontal growth of predominantly monolayer MoS$_2$ ribbons? Here, we present our qualitative understanding of the growth processes and mechanism based on our experimental observations as schematically summarized in Figure 5. The precursor for VLS growth is the formed liquid droplets (Figure 5a) via reaction between the vapors of MoO$_3$ and NaCl as discussed above. When Na$_2$MoO$_4$ is used, liquid droplets form by direct melting (Supplementary Fig. 7). Subsequently, vaporized sulfur dissolves into the liquid droplet until saturation, triggering the formation of a MoS$_2$ nucleus at the droplet-substrate interface (Figure 5b). Continuous precipitation of MoS$_2$ allows commensurate growth on the support crystal structure (Figure 5c). As the MoS$_2$ ribbon growth proceeds, the liquid droplet begins to crawl on the substrate surface (Figure 5d). The origin of the crawling mode has been discussed in the literature but a unified theory has not been established due to complex kinetics of the processes involved[43,44]. Here, we propose that the driving force for liquid droplet motion results primarily from the competition between the interface free energies of the droplet-MoS$_2$ ribbon ($\sigma_1$, corresponding area $A_1$) and droplet-substrate ($\sigma_2$, corresponding area $A_2$) (Figure 5e). With the condition of $\sigma_1 > \sigma_2$, the total interface free energy ($G_{interface}$) keeps increasing ($\delta G_{interface} = (\sigma_1 - \sigma_2)\delta A_1 > 0$) as the MoS$_2$ nucleation and growth at the droplet-substrate interface (Figure 5b and c). Then, the droplet de-wets MoS$_2$ ribbon form "Position X" of Figure 5c to minimize the $G_{interface}$ (shrinkage of $A_1$). As a result, a quasi-equilibrium state would be achieved as illustrated in Figure 5f. While the newly-grown MoS$_2$ ribbon further increases the area of higher-energy droplet-MoS$_2$ ribbon interface ($\sigma_1$) by $\delta A_1$ which means the $G_{interface}$ keeps increasing. The increased $G_{interface}$ leads to a driving force $F$ for the droplet to displace laterally and restore the quasi-equilibrium state at the tip of the ribbon (Figure 5g). This is similar to the spontaneous lateral motion of a water droplet from a low to higher surface energy regions of the surface[45]. While the magnitude of the interfacial energies is not known, finite difference in the surface energy of the ribbon and the substrate suggests that the above scenario is possible. The surface energy of NaCl is significantly greater than that of monolayer MoS$_2$[46,47], and similarly, monolayer MoS$_2$ on SiO$_2$/Si substrate is expected to exhibit higher surface energy compared to bilayer MoS$_2$[47]. The different surface tensions of MoS$_2$ ribbon and growth substrates are well satisfied the condition of $\sigma_1 > \sigma_2$ which makes the motion of droplets possible during the continuous precipitation of MoS$_2$ from droplets. Other forces such as capillary and Marangoni forces may also play a role in the lateral motion[48]. As the droplet crawls on the surface, sulfur and also possibly MoO$_3$ continue to dissolve into the droplet inducing further continuous growth yielding the observed ribbon structures. The ripple-like features and local strains in the ribbons most likely resulted from the compressive stress induced by the droplet during growth. The predominance of monolayer growth may be explained by the fast motion of the liquid droplet, which allows limited time for the nucleation and growth of the second layer. Due to the finite vapor pressure of $MoO_3$ and S during growth, the ribbons can also grow at the edge of ribbons via vapor-solid (VS) or VSS conversion of the precursors. The strain-free edge region of the ribbons most likely resulted from such non-VLS growth (Figure 2d and Figure 5d).

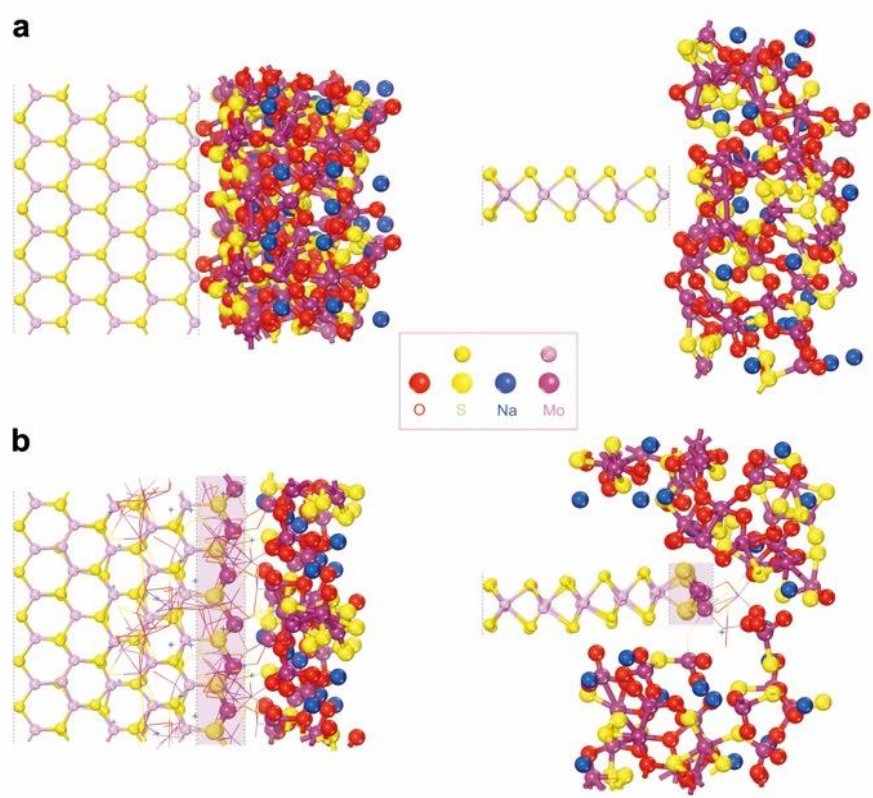

**Figure 6 | Density functional theory based molecular dynamic (DFT-MD) simulation of $MoS_2$ precipitation process.** (a) Initial state of the junction between S-rich Na-M-O droplet and adjacent $MoS_2$ nanoribbon. The left and right panels show the top and side view of the junction. (b) Mo and S atoms attach on the edge of $MoS_2$ nanoribbon, nearly completing a new row of atoms after annealing from 1500 to 1600 K for 17.5 ps (highlighted with a colored box).

In order to gain insight into the liquid-solid transformation, we performed density functional theory based molecular dynamic (DFT-MD) simulations of the precipitation process[49-52].

Initially, 22 Mo atoms and 43 S atoms were dissolved into a disordered $Na_{21}Mo_{21}O_{69}$ droplet, which was placed next to a $MoS_2$ nanoribbon with its Mo-terminated zigzag edge exposed to the droplet. The system was annealed from 1500 to 1600 K (see Method) in a period of ~ 17.5 ps. The initial and final structures are shown in Figure 6 (see Movie S1 and S2, Supplementary Information for the full animation). The simulation shows attachment of S and Mo to the zigzag edge, nearly completing a new row of $MoS_2$. It is noteworthy that $MoS_2$ is not oxidized despite the presence of large number oxygen atoms. We also observe nucleation of $MoS_2$ clusters in regions that are rich of Mo/S atoms, further supporting the feasibility of liquid-mediated nucleation and growth of $MoS_2$.

**Conclusions**

In summary, we have demonstrated the VLS growth of monolayer $MoS_2$ nano- and micro-ribbons on NaCl crystals and VLS-based homoepitaxy on monolayer $MoS_2$ using salt-assisted CVD. The VLS growth is made possible by liquid solution formation due to the vapor phase reaction between $MoO_3$ and NaCl. While there are some similarities with conventional VLS growth of nanowires and nanotubes, there are notable differences: the growth is strictly limited to the interface between the liquid phase precursor and the substrate surface; it yields predominantly monolayer products; and it is a van der Waals epitaxy. We envision that optimized VLS growth of 2D TMDs will allow rapid synthesis of nano- and micro-ribbon arrays and complex heterostructures that can be readily integrated into nanoelectronic and photonic devices. By identifying a suitable liquid phase intermediate compound, we believe that it will be possible to realize the direct 1D growth of a range of van der Waals layered materials. Our findings also open up avenues for the fundamental studies of lateral quantum confinement and dimensional crossovers in these 2D materials.

**Methods**

**Salt-assisted VLS growth of MoS$_2$ ribbons:** (1) For growing MoS$_2$ ribbons on NaCl crystals, a piece of freshly cleaved NaCl was placed above the MoO$_3$ powder on an alumina crucible. (2) For homoepitaxial ribbon growth, a SiO$_2$/Si substrate with pre-grown monolayer MoS$_2$ was loaded near the NaCl crystal. Sulfur was loaded in the low temperature upstream zone that reached ~230 °C during growth. 100 sccm Ar was used as carrier gas. The growth was conducted at 700-730 °C for 2-5 min. (See Supplementary Information for details)

**STEM analysis of MoS$_2$ ribbons:** STEM imaging was performed by using JEOL 2100F microscope equipped with a cold field emission gun operating at 60 kV and dodecaple correctors. The probe current is about 20-25 pA. The convergence semi-angle and the inner acquisition semi-angle are 35 mrad and 79 mrad.

**SHG microscopy of homoepitaxially grown ribbons**: An 800-nm femtosecond (fs) laser light with a repetition rate of 80 MHz and a duration of 140 fs delivered by a Ti: Sapphire oscillator (Coherent, Chameleon Ultra Ⅱ) was focused on the MoS$_2$ samples using 100 air objective lens (NA = 0.90) of an inverted microscope (Nikon). The nonlinear optical signals generated by the MoS$_2$ samples were obtained by using the same objective lens and directed to a spectrometer (Princeton Instruments) for SHG spectroscopy or to a photomultiplier tube (PicoQuant, PMA 182) for SHG mapping. During mapping, the sample was raster scanned by a piezo-actuated 3D nano-positioning stage (Physik Instrumente).

**DFT-MD calculation**: The DFT-MD simulation were performed with the Vienna Ab Initio Simulation Package (VASP)[49]. The exchange-correlation functional was described by the Perdew-Burke-Ernzerhof version of generalized gradient approximation[50] and the core region by the projector augmented wave method[51]. Considering the time-consuming nature of the DFT-MD, the plane-wave cutoff is set to be 300 eV and the Brillouin zone is sampled at Γ point only. All DFT-MD trajectories were run with the canonical ensemble (NVT) using a Nosé thermostat[52] at the temperature from 1500 K (~ 6 ps), 1400 K (~ 3 ps) to 1600 K (~ 8.5 ps), which is higher than the real experimental temperature (~ 1000 K) but is necessary to accelerate the atomic evolution to observe a full annealing process in a short simulation time. The time step of the DFT-MD simulation is 2 fs. The total time scales of the trajectories are more than 17.5 ps. The unit cell used in the simulation has vacuum layer in the growth direction to avoid interaction between adjacent images and is periodic in another two directions. In the lattice used in simulations, initially the nanoribbon has three hexagons in the lateral direction and four hexagons in the VLS-growth direction. The freestanding end of the nanoribbon is fixed during the simulation.

**Acknowledgements**

G.E. acknowledges Singapore National Research Foundation for funding the research under NRF Research Fellowship (NRF-NRFF2011-02) and medium-sized centre programme. G.E. also acknowledges support by Ministry of Education (MOE), Singapore, under AcRF Tier 2 (MOE2015-T2-2-123, MOE2017-T2-1-134). Y.L. and K.S. acknowledge the support from JSPS KAKENHI (JP16H06333). J. W. acknowledges the support from A*STAR IMRE/15-2C0111. F. D. acknowledges the support from the Institute for Basic Science (IBS-R019-D1). S.L. acknowledges Yang Sun for helpful discussion.

**Author contributions**

S.L. designed and conducted the VLS growth. Y.L. performed the STEM characterization of $MoS_2$ ribbons. S.L., J.W., Y.S., W.Z., F.D. and G.E. interpreted the VLS growth of $MoS_2$ ribbons. Z.W. and H.Z. performed the SHG mapping. S.L., J.W. and L.C. studied the Raman, PL, AFM and electrical properties of $MoS_2$ ribbons. S.L. and Q.Z. performed the TGA and XRD experiments. S.L. and Z.H. conducted the growth of $MoX_2$, $WX_2$ (X=S, Se, Te) from sodium molybdate and sodium tungstate. S.L., Y.L., W.Z., F. D. and G.E. wrote the paper. All the authors discussed and commented on the manuscript.

**Competing financial interests**

The authors declare no competing financial interests.

Supplementary Information for

# Vapor-Liquid-Solid Growth of Monolayer MoS$_2$ Nanoribbons

*Shisheng Li, Yung-Chang Lin, Wen Zhao, Jing Wu, Zhuo Wang, Zehua Hu, Youde Shen, Dai-Ming Tang, Junyong Wang, Qi Zhang, Hai Zhu, Leiqiang Chu, Weijie Zhao, Chang Liu, Zhipei Sun, Takaaki Taniguchi, Minoru Osada, Wei Chen, Qing-Hua Xu, Andrew Thye Shen Wee, Kazu Suenaga, Feng Ding and Goki Eda*

Corresponding authors: g.eda@nus.edu.sg (G.E); shishengli1108@gmail.com (S.L.)

***Chemical vapor deposition (CVD) of MoS$_2$ nanoribbons on NaCl crystals:*** The growth was conducted in ambient pressure with a movable tube furnace (MTI OTF-1200X, 20-mm-long heating zone) that allows rapid heating and cooling as illustrated in Supplementary Fig. 1a (Fast-heating CVD). A piece of freshly cleaved NaCl crystal (10×10×1 mm$^3$, MTI) was placed on an alumina crucible (77/12/8 mm, L/W/H) with the cleaved surface faced down above MoO$_3$ powder (~5-10 mg, Alfa-Aesar) as illustrated in Supplementary Fig. 1b(I). The environmental humidity was kept lower than 50% to keep the NaCl surface dry without deliquescence. Sulfur (~30 mg) was loaded in the upstream region outside the heating zone (position 1, Supplementary Fig. 1a) during the temperature ramping process to keep it from evaporation. For a typical growth, 100 sccm Ar (99.995%) was pass through a 1000-mm-long and 50-mm-diameter (O.D.) quartz tube. The furnace temperature ramping rate is 25 $^\circ$C/min. After the tube furnace reached the growth temperature (700-730 $^\circ$C) and stabilized for ~2 min, the furnace was shifted to start the vaporization of sulfur and MoO$_3$ powder with the substrate at the centre of heating zone. The growth was terminated after ~2-5 min by moving the precursors and the substrate out of the heating zone and rapidly cooling to room temperature.

***CVD of MoS₂ ribbons on monolayer MoS₂ film:*** For the epitaxial VLS growth of $MoS_2$ ribbons on monolayer $MoS_2$ film, a piece of NaCl crystal (10×10×1 mm$^3$, MTI) and a piece of $SiO_2$/Si substrate with as-grown monolayer $MoS_2$ film were loaded on an alumina crucible as shown in Supplementary Fig. 1b(II). ~5-10 mg $MoO_3$ and ~30 mg sulfur (position 1, Supplementary Fig. 1a) were used as the source materials. Furthermore, ~10 mg sulfur (position 2, Supplementary Fig. 1a) was evaporated during the temperature ramping process to prevent the monolayer $MoS_2$ film form being oxidized. The optimized VLS epitaxial growth of $MoS_2$ ribbons on monolayer $MoS_2$ film was achieved with following recipe: Fast-heating CVD, ambient pressure, 100 sccm Ar (99.995 %), temperature ramping rate of 25 °C/min and hold at 700-730 °C for ~5 min (~2 min for stabilization, ~3 min for growth). KCl also works well as growth precursor as NaCl. In the case of KCl-assisted growth, the recipe is almost the same as the case of NaCl, and the optimal growth temperature is 675-700 °C.

***Growth and characterization of monolayer MoS₂ film:*** ~0.5 ml of 0.2 mg/ml of NaCl aqueous solution was dropped in an alumina crucible (L/W/H: 77/12/8 mm). The crucible was baked on a hot plate at 120 °C until dry. Then, ~1 ml of 5 mg/ml $MoO_3$ aqueous suspension was dropped in the crucible and dried at 120 °C. A piece of $SiO_2$/Si substrate (NOVA), typical size of 20-25/12.5mm (L/W), was loaded on the crucible with polished face down. ~30 mg sulfur was loaded in the upstream, ~120 mm away from the center of the furnace. The fast-heating CVD was used for the growth of $MoS_2$ film. 100 sccm Ar (99.995%) was used as carrier gas. During the temperature ramping process (25 °C/min), the sulfur and crucible were kept at the low temperature. The center of the crucible is ~60 mm away from the center of the furnace. After the furnace reached the growth temperature (700 °C) and stabilized for ~2 min, the furnace was moved to heat the sulfur and crucible to initiate the growth of monolayer $MoS_2$ film for ~3 min. Finally, the growth was terminated by moving the precursors and the substrate out of the heating zone.

***Sublimation of NaCl:***
A piece of NaCl crystal (10×10×1 mm$^3$, MTI) and a piece of $SiO_2$/Si substrate (NOVA) were loaded on an alumina crucible as shown in Supplementary Fig. 4a. The sublimation of NaCl crystal was performed at 700 °C in Ar for 10 min. Optical image shows a gradient of small NaCl crystals deposited on the $SiO_2$/Si substrate, suggesting sublimation of NaCl indeed occurs below its melting point (801 °C). The surface of SiO$_2$/Si substrate was checked by SEM and EDX (Supplementary Fig. 4b-d).

*Co-sublimation of NaCl and MoO$_3$*

The sources, NaCl and MoO$_3$, were separated as shown in Supplementary Fig. 4e. The MoO$_3$ powder was loaded in the crucible under a piece of SiO$_2$/Si substrate. After annealing at 700 °C in Ar for 10 min, the surface of the SiO$_2$/Si substrate was checked by SEM and EDX (Supplementary Fig. 4f-h). "Liquid droplets" with elements of Na, Mo, O and Cl were observed indicating efficient reaction of NaCl and MoO$_3$.

*Thermogravimetric (TG) study of pure MoO$_3$ and mixed MoO$_3$-NaCl:* 11.64 mg MoO$_3$ powder and 34.46 mg mixed MoO$_3$-NaCl (20 wt%) were loaded in alumina crucibles for TG experiments, respectively. The TG experiments were carried out in N$_2$ atmosphere with a temperature ramping rate of 10 °C/min to 1000 °C. The TG-DTG data and optical images of the crucibles after the TG experiments are showing in Supplementary Fig. 5a and b.

*In-situ observation of the melting of mixed MoO$_3$-NaCl:* ~60 mg mixed MoO$_3$-NaCl was loaded in a clean alumina crucible. Typical CVD process without sulfur was performed. Photos were taken at room temperature (R.T.), 400 °C, 530 °C, 600 °C and 700 °C, respectively (Supplementary Fig. 5c). The white alumina crucible is "stained" at 530 °C indicating the release of vapor phase product. This matches well with our TG-DTG data. At 600 °C and above, the melting of the reaction products and wetting of the bottom of alumina crucible were directly observed (Supplementary Fig. 5c).

*X-ray diffraction (XRD) of the reaction product of mixed MoO$_3$-NaCl:* To prepare XRD sample, ~200 mg mixed MoO$_3$-NaCl was loaded in a clean alumina crucible. After annealing at 700 °C in Ar for 20 min. The reaction product was collected for XRD (Supplementary Fig. 5d). XRD pattern shows that reaction product is a mixture of Na$_2$Mo$_2$O$_7$ and residual NaCl.

*Regrowth of MoS$_2$ from the reaction product of mixed MoO$_3$-NaCl:* The obtained mixture of Na$_2$Mo$_2$O$_7$ and residual NaCl was dispersed on a SiO$_2$/Si substrate for growing MoS$_2$. The as-grown MoS$_2$ shows a morphology of "coffee-ring" indicating the melting of the Na$_2$Mo$_2$O$_7$ during the formation of MoS$_2$ (Supplementary Fig. 5e-g).

***The chemical composition of the liquid droplets for VLS growth:*** For the NaCl-assisted growth of MoS$_2$ ribbons on monolayer MoS$_2$ at 700 °C, the growth was terminated within 1 min to preserve the liquid droplets at the end of ribbons (Supplementary Fig. 6a-f). The EDX spectra of these liquid droplets reveal the existence of elements of Mo, S, Na, O and trace amount of Cl which matches well with our XRD data (Supplementary Fig. 5d). Similarly, the slow growth of MoS$_2$ ribbons was also achieved with KCl as precursor at a lower growth temperature (675 °C). The trace at the tails of the MoS$_2$ ribbons clearly demonstrate the melting of the particles (Supplementary Fig. 6g and h). EDX spectra of the particles at the heads confirmed the existence of elements of Mo, S, K, O and trace amount of Cl (Supplementary Fig. 6i).

Thus, we conclude that the liquid droplets for growing MoS$_2$ ribbons is mainly a compound of Na/K, Mo and O. Cl may also contribute in lowering the melting point of the mixture. However, it is likely that Cl is lost quickly in gas phase MoO$_2$Cl$_2$[1] during growth.

***Possible chemical reaction route:*** Based on the above analysis, we deduced the possible VLS growth process as following. During the temperature ramping to the growth temperature of 700-730 °C, the co-sublimated MoO$_3$ and NaCl first react as equation (1)[1].

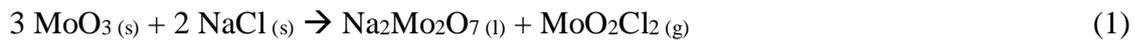
3 MoO$_3$ $_{(s)}$ + 2 NaCl $_{(s)}$ → Na$_2$Mo$_2$O$_7$ $_{(l)}$ + MoO$_2$Cl$_2$ $_{(g)}$ (1)

Then, sulfur vapor dissolved into the liquid-phase Na$_2$Mo$_2$O$_7$. Finally, monolayer MoS$_2$ nanoribbons were grown on the substrate.

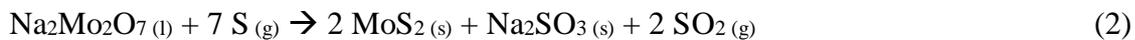
Na$_2$Mo$_2$O$_7$ $_{(l)}$ + 7 S $_{(g)}$ → 2 MoS$_2$ $_{(s)}$ + Na$_2$SO$_3$ $_{(s)}$ + 2 SO$_2$ $_{(g)}$ (2)

***Growth of MoS$_2$ ribbons with patterned Na$_2$MoO$_4$ precursor on sapphire substrates:*** To prepare patterned Na$_2$MoO$_4$, sapphire substrates with line patterns on photoresist were fabricated by photolithography. Then, ~1 ml of 20 mg/ml Na$_2$MoO$_4$ aqueous solution was spin-coated on the sapphire substrates. A standard lift-off process was employed to remove photoresist. The as-fabricated substrates with Na$_2$MoO$_4$ line patterns were used for the growth of MoS$_2$ ribbons. The growth was done at 700 °C in 100 sccm Ar for ~5 min (Supplementary Fig. 7)

***VLS growth of monolayer MoX$_2$ and WX$_2$ (X = S, Se, Te) crystals from Na$_2$MoO$_4$ and Na$_2$WO$_4$ precursors:*** For the growth of MoX$_2$ (X = S, Se), 2 mg/ml Na$_2$MoO$_4$ (T$_m$ = 687 °C) aqueous solution was spin-coated on sapphire (Namiki Precision) or SiO$_2$/Si substrates (Nova). The optimized growth condition for MoX$_2$ (X = S, Se) is 700-725 °C in Ar for ~5 min. For the growth of MoTe$_2$ and WX$_2$ (X = S, Se, Te), 2 mg/ml NaMoO$_4$ or Na$_2$WO$_4$ (T$_m$ = 698 °C) aqueous solution was spin-coated on sapphire or SiO$_2$/Si substrates, respectively. The optimized growth condition for MoTe$_2$ and WX$_2$ (X = S, Se, Te) is 725-775 °C in Ar/H$_2$ (5%) for ~5 min. For the growth of MoS$_2$/MoSe$_2$ heterostructure, CVD-grown monolayer MoS$_2$ crystals were wet-transferred on a new SiO$_2$/Si substrate first. Then, 2 mg/ml Na$_2$MoO$_4$ aqueous solution was spin-coated on the SiO$_2$/Si substrate. MoSe$_2$ monolayers were grown at 725 °C for MoS$_2$/MoSe$_2$ heterostructure. Supplementary Fig. 8c-j show the morphology of monolayer MoX$_2$, WX$_2$ (X = S, Se, Te) crystals and MoS$_2$/MoSe$_2$ heterostructure grown with Na$_2$MoO$_4$ and Na$_2$WO$_4$ precursors, respectively. All the growth was conducted in a 2-inch (O.D.) tube furnace with a total flow rate (Ar, Ar/H$_2$) of 100 sccm.

***Large-area growth of monolayer MoS$_2$ ribbons with Na$_2$MoO$_4$ precursor on sapphire substrates:*** For the growth of MoS$_2$ ribbons, ~2 ml of 4 mg/ml Na$_2$MoO$_4$ aqueous solution was spin-coated on a 2-inch sapphire wafer (Namiki Precision). The growth was conducted in a 2-inch tube furnace. The optimized growth condition is at ~700-775 °C in 100 sccm Ar for ~2-5 min (Supplementary Fig. 8k and l)

***Large-area uniform growth of monolayer MoS$_2$ film with Na$_2$MoO$_4$ precursor on sapphire substrates:*** For the growth of monolayer MoS$_2$ film, ~4 ml of 20 mg/ml Na$_2$MoO$_4$ aqueous solution was spin-coated on a 2-inch sapphire wafer. The growth was conducted in a 3-inch tube furnace. The optimized growth condition is at 725 °C in 200 sccm Ar for ~5 min (Supplementary Fig. 8m and n).

***Wet transfer of monolayer MoS$_2$ nanoribbons from NaCl crystal:*** A layer of PMMA film was spin-coated on the NaCl crystal with as-grown monolayer MoS$_2$ nanoribbons (PMMA solution: 950K-5%, 3000 rpm, 60 s). The NaCl crystal was removed by simply dissolving it in DI water.

*Field effect transistor fabrication and measurement:* Standard e-beam lithography patterning and lift-off procedures were preformed to define the electrodes. 2 nm Cr and 50 nm Au film was deposited as electrodes by thermal evaporator. Electrical measurements were conducted in $N_2$-filled glovebox at room temperature. The applied source-drain bias ($V_{ds}$) is 1 V. All the transport curves are presented in effective 2D electrical conductivity, $\sigma = (I_{ds}/V_{ds})*(L/W)$, where the $I_{ds}$ is source-drain current, $L$ and $W$ are the channel length and width, respectively. From the linear regime of the transfer curves, the filed-effect mobility was calculated using the equation $\mu = (1/C_{ox})(d\sigma/dV_{bg})$, where $C_{ox}$ is the area capacitance of the gate oxide ($1.2 \times 10^{-4}$ F/m$^2$ for 285 nm $SiO_2$), $V_{bg}$ is the back-gate bias.

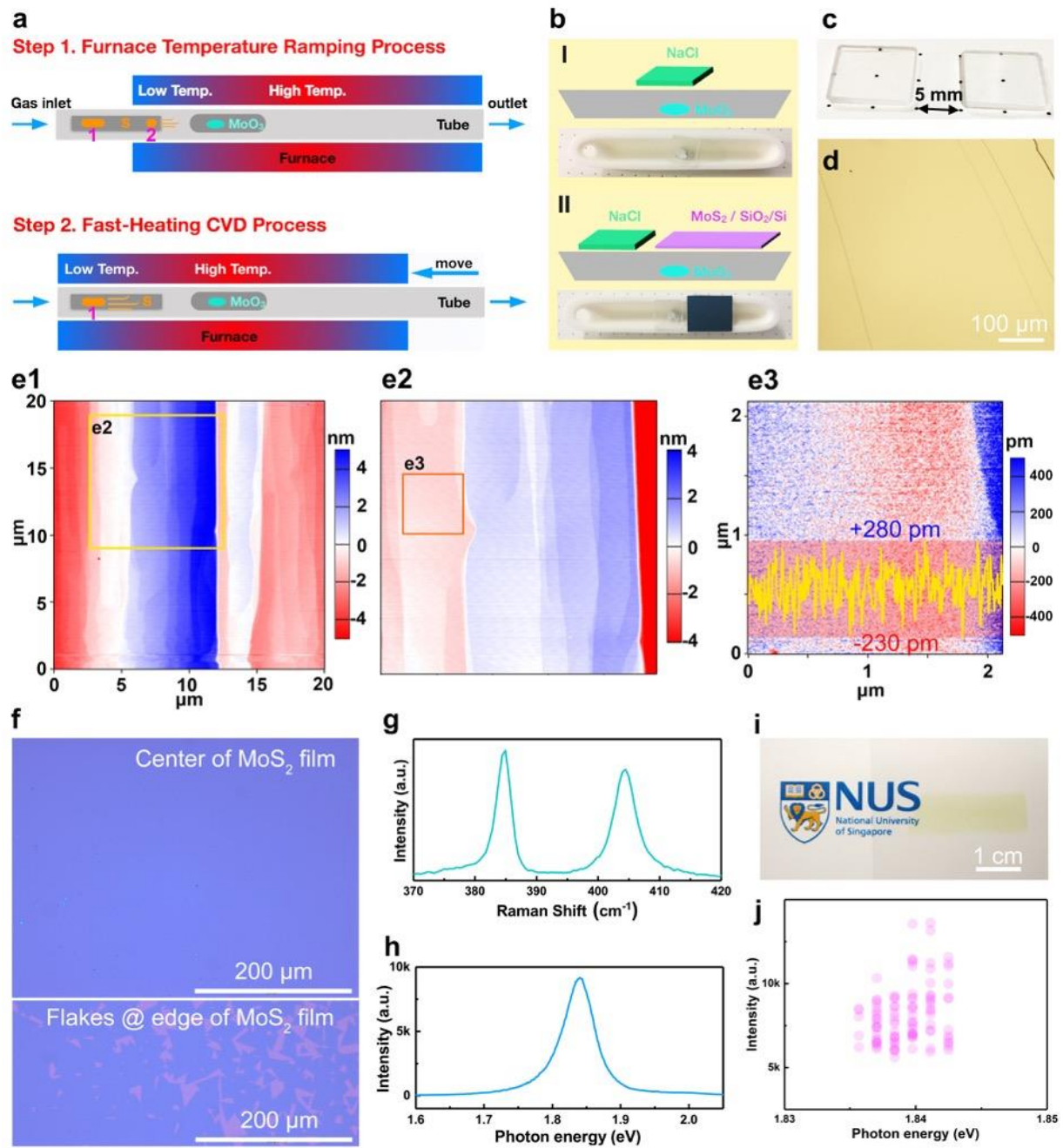

**Supplementary Figure 1 | Schematic illustration of fast-heating CVD process and characterization of the raw materials for VLS growth of MoS$_2$ ribbons.** (a) Schematic illustration of the fast-heating CVD process. (b) Schematic illustrations and optical images showing the configurations for growth of MoS$_2$ ribbons on (I) NaCl crystal and (II) monolayer MoS$_2$ film. (c) Photograph of two freshly cleaved NaCl crystals. (d) Optical image of the surface of a freshly cleaved NaCl crystal. (e1-e3) AFM images showing the surface of NaCl crystal with parallel surface steps. The flat surface has a roughness of approximately 0.5 nm. (f) Optical images of monolayer MoS$_2$ film and flakes grown on a SiO$_2$/Si substrate used

for the homoepitaxial growth of MoS$_2$ ribbons. (g, h) Typical Raman and PL spectra of monolayer MoS$_2$ film. (i) Optical image of a transferred monolayer MoS$_2$ film on PET substrates. (j) The PL peak energy and intensity distribution of a monolayer MoS$_2$ film. 100 PL spectra were collected from the whole surface of a monolayer MoS$_2$ film (~2.5×1 cm$^2$).

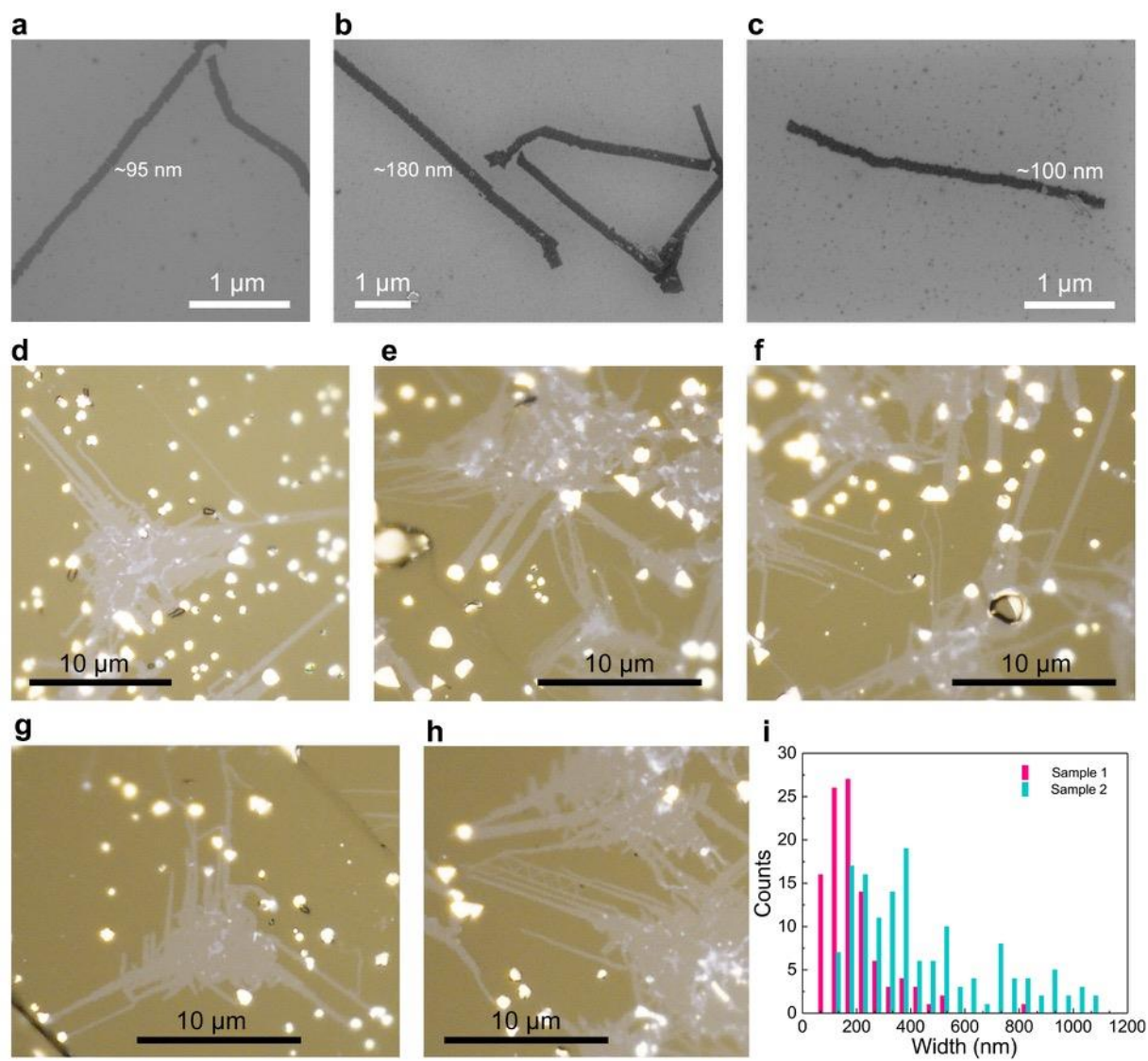

**Supplementary Figure 2 | Width distribution of MoS$_2$ nanoribbons grown on NaCl crystal.** (a-c) SEM and (d-h) optical images of MoS$_2$ nanoribbon structures. (i) Histogram showing the width distribution of >100 nanoribbons in two samples. (a-c) MoS$_2$ nanoribbons grown with ~1 mg MoO$_3$ powder, the average nanoribbon width is ~195 nm. (d-h) MoS$_2$ nanoribbons grown with ~20 mg MoO$_3$ powder, the average nanoribbon width is ~440 nm.

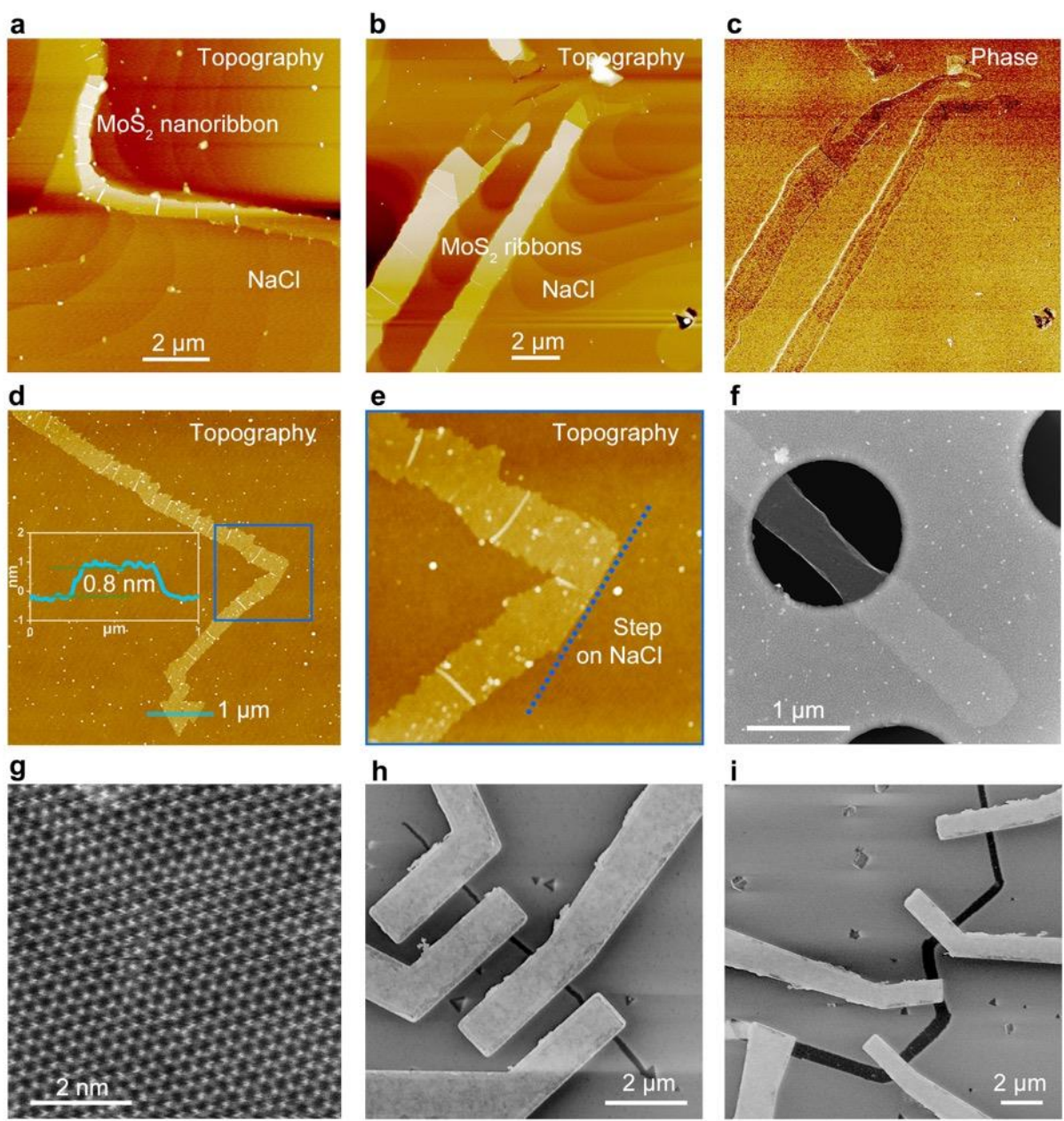

**Supplementary Figure 3 | Morphology of as-grown and transferred monolayer MoS$_2$ ribbons.** (a, b) AFM topography images of MoS$_2$ ribbons grown on NaCl crystal. (c) Corresponding phase image of (b). (d) AFM topography image of a transferred MoS$_2$ nanoribbon on SiO$_2$/Si substrate. (e) AFM topography image of a MoS$_2$ nanoribbon showing the details around the kink that correspond to the step on NaCl crystal surface. This morphology suggests that the migration of the liquid droplet was obstructed by the NaCl surface step and the growth direction was changed therefore. (f) A TEM image showing a general view of a MoS$_2$ nanoribbon on a TEM grid. (g) A high-resolution STEM image of MoS$_2$ nanoribbon indicating its high crystallinity. (h, i) SEM images of (h) straight and (i)

kinked MoS$_2$ nanoribbon devices. The width of the narrowest segment of the straight nanoribbon is ~60 nm.

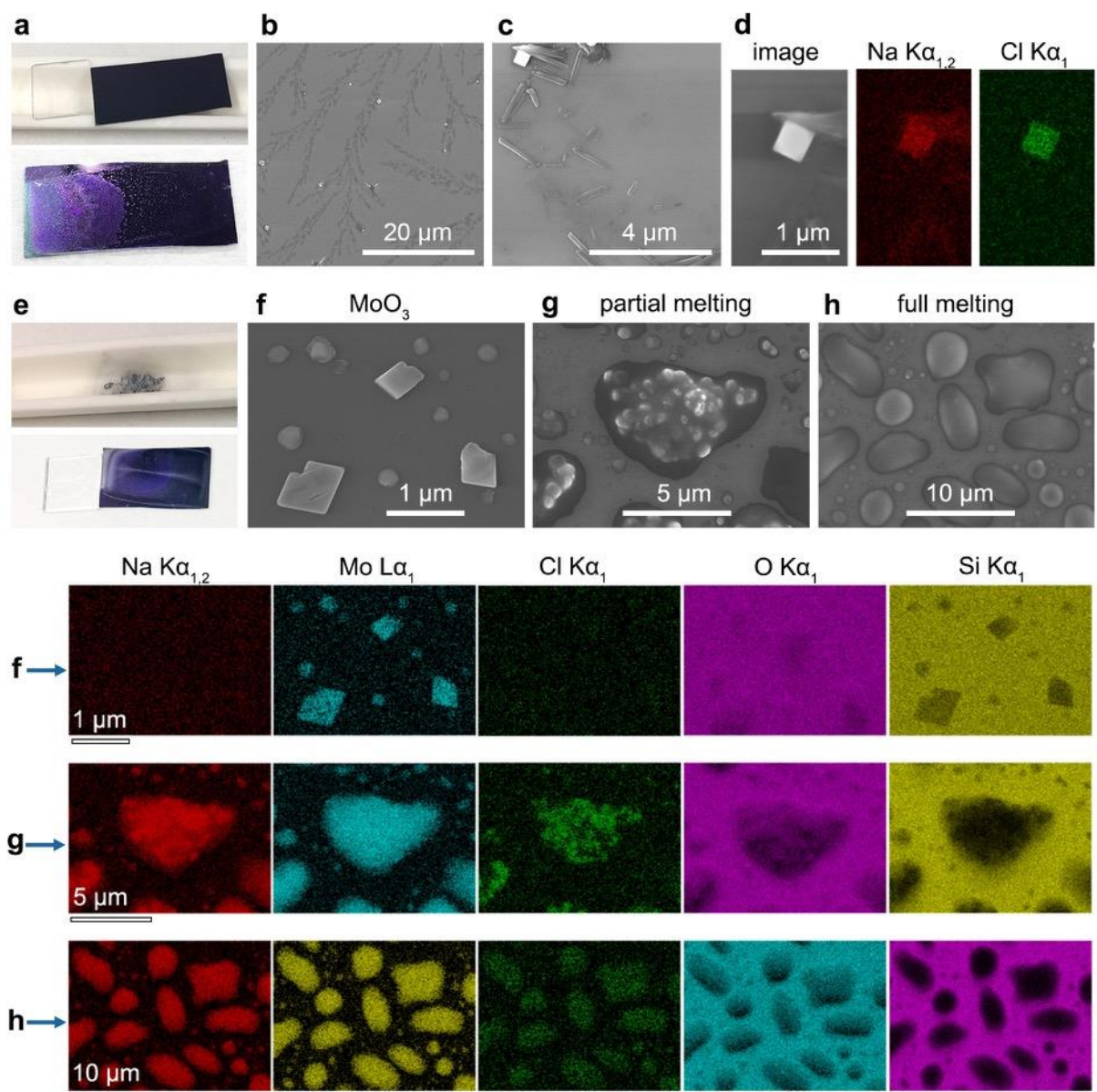

**Supplementary Figure 4 | Sublimation of NaCl and co-sublimation of NaCl and MoO$_3$.** (a) Optical image of the sublimed NaCl on a clean SiO$_2$/Si substrate at 700 °C in Ar. (b, c) SEM images of NaCl crystals deposited on the SiO$_2$/Si substrate. (d) SEM image and corresponding EDX mapping images of Na K$\alpha_{1,2}$ and Cl K$\alpha_1$. (e) Optical image of co-sublimated NaCl and MoO$_3$ on a clean SiO$_2$/Si substrate at 700 °C in Ar. (f-h) SEM images and corresponding EDX mapping images (Na K$\alpha_{1,2}$, Mo L$\alpha_1$, Cl K$\alpha_1$, O K$\alpha_1$ and Si K$\alpha_1$) of (f) MoO$_3$, (g) partially and (h) fully liquefied reaction products of mixed MoO$_3$-NaCl,

respectively. Samples shown in (g) and (h) contain low-melting-point $Na_2Mo_2O_7$ and residual NaCl.

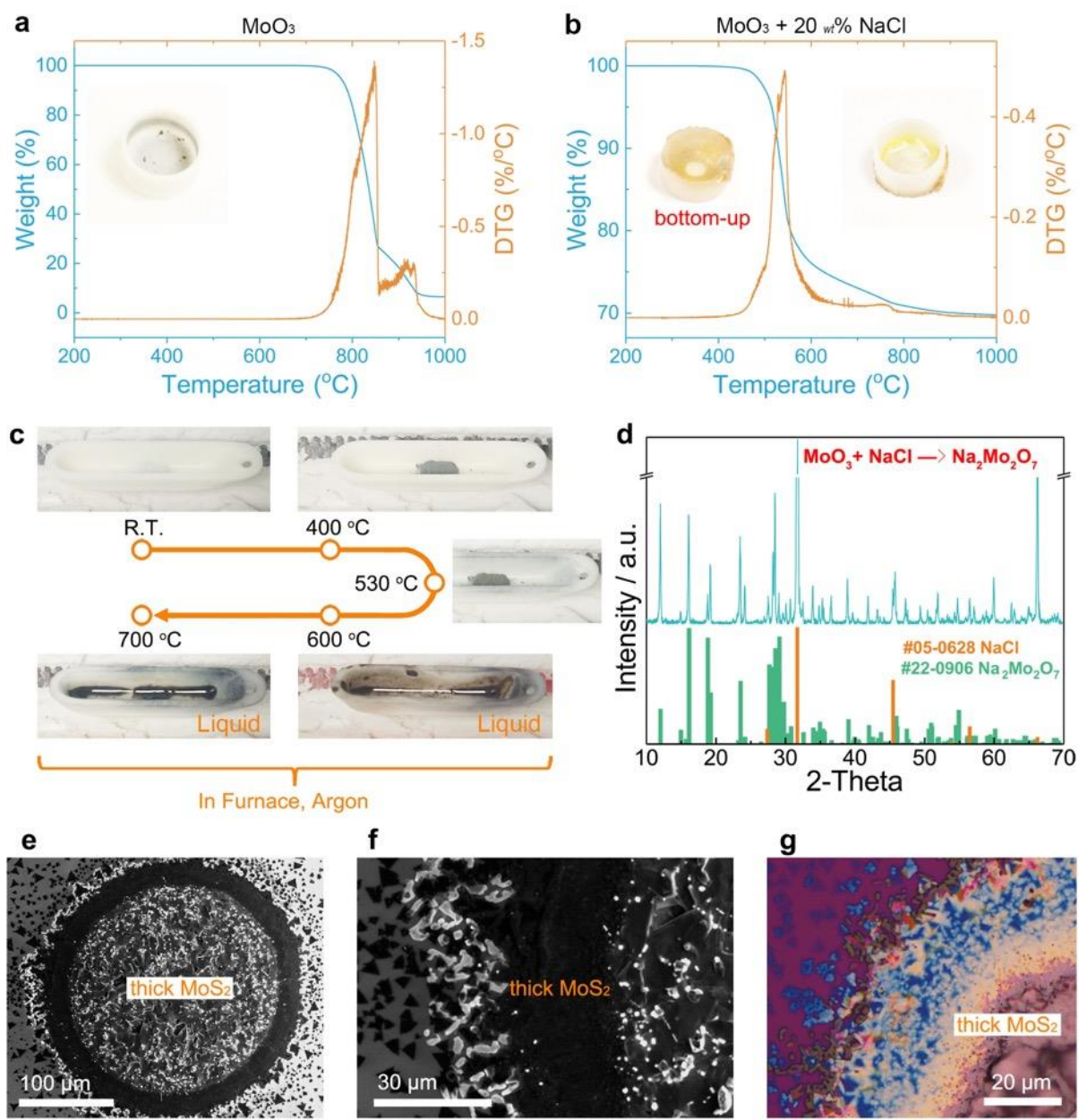

**Supplementary Figure 5 | Reaction of MoO$_3$-NaCl mixture.** (a, b) TG-DTG curves of (a) MoO$_3$ and (b) mixed MoO$_3$-NaCl in N$_2$ with a temperature ramping rate of 10 °C/min. Insets are optical images showing the residuals in and out alumina crucibles after TG study. (c) Optical images showing melting of MoO$_3$-NaCl mixture when heated to 600 °C and above in Ar. (d) XRD pattern of the reaction product of MoO$_3$-NaCl mixture at 700 °C. It contains Na$_2$Mo$_2$O$_7$ and residual NaCl. (e, f) SEM and (g) optical images of MoS$_2$ obtained by

sulfurization of $Na_2Mo_2O_7$. The "coffee-ring" pattern suggests that the precursors were in a liquid state during the growth.

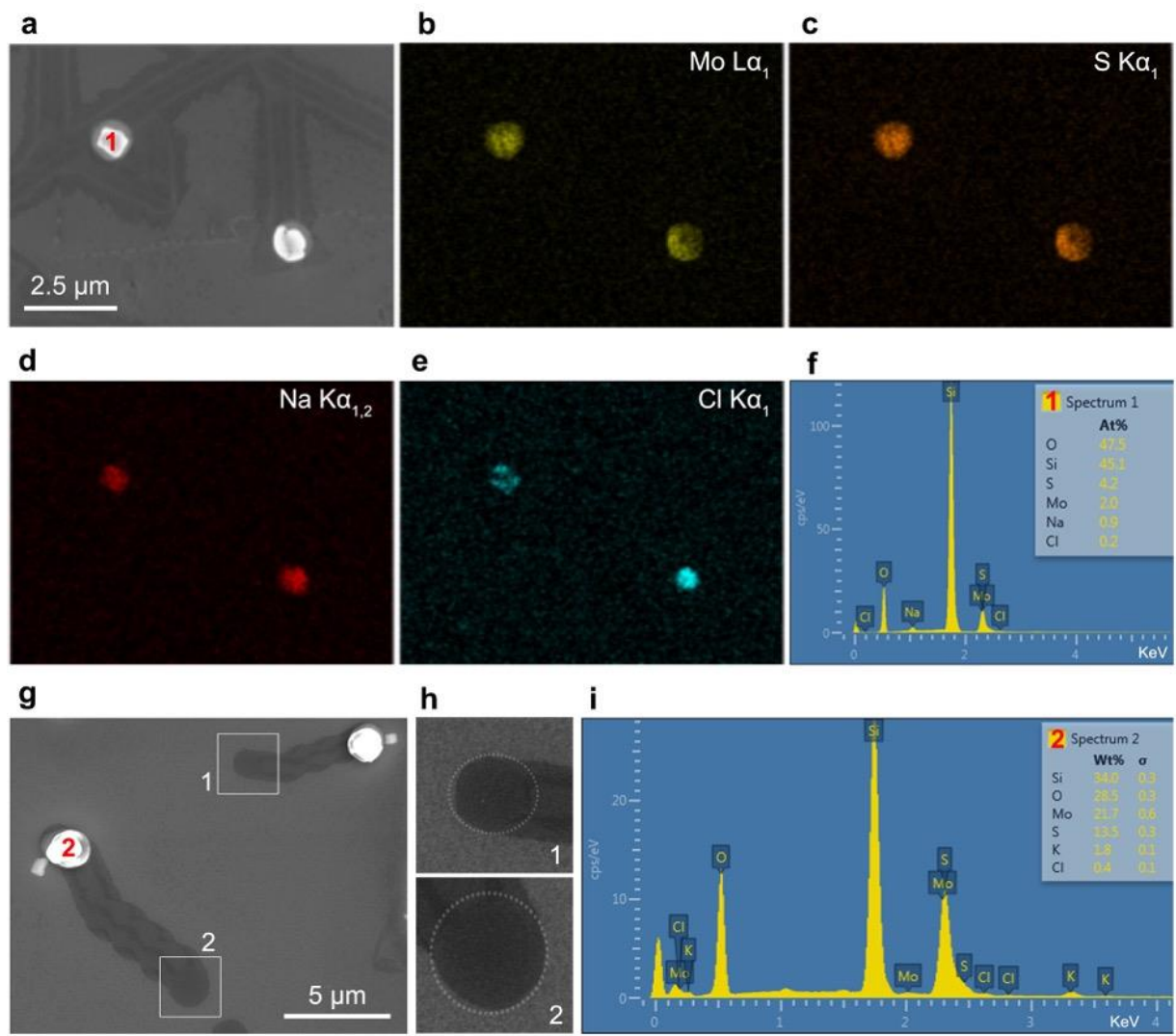

**Supplementary Figure 6 | SEM/EDX analysis of the particles terminating MoS$_2$ ribbons.** (a) SEM image of two particles terminating two MoS$_2$ ribbons grown on monolayer MoS$_2$ film. (b-e) Corresponding EDX element mapping images of (a): (b) Mo L$\alpha_1$, (c) S K$\alpha_1$, (d) Na K$\alpha_{1,2}$ and (e) Cl K$\alpha_1$. (f) EDX spectrum of the particle 1. (g-i) KCl-assisted growth of MoS$_2$ ribbons on monolayer MoS$_2$ film (675 °C). (g) SEM image shows the heads and tails of two short MoS$_2$ ribbons. (h) The morphology of the tail region the ribbons (the end that is not terminated by a particle). The round feature is indicative of the liquid-state of the particle during the growth. (i) EDX spectrum of the particle 2 which contains elements of Mo, S, K, O and small amount of Cl.

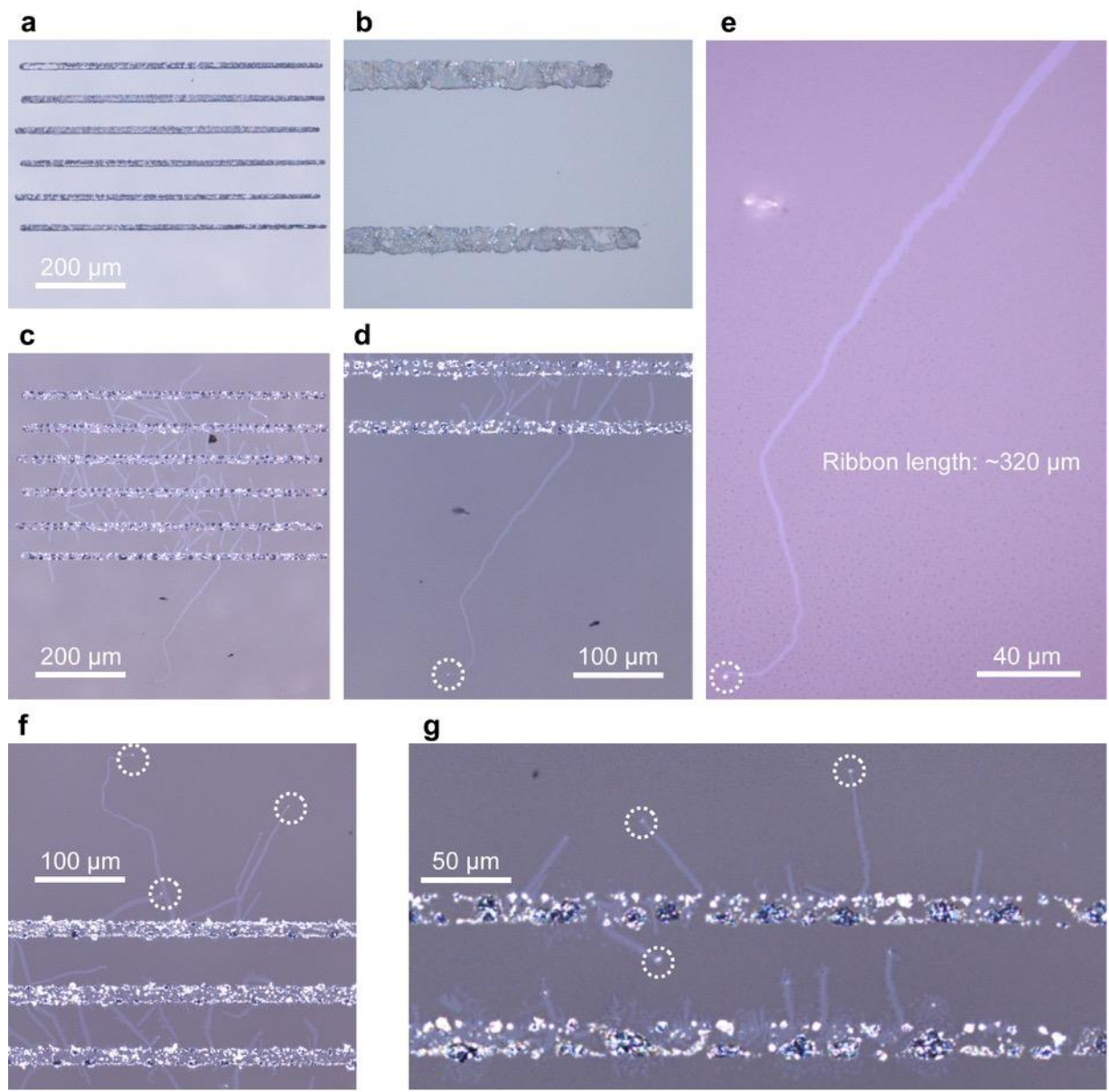

**Supplementary Figure 7 | VLS growth of MoS$_2$ ribbons with patterned Na$_2$MoO$_4$ precursor.** (a, b) Optical images of patterned Na$_2$MoO$_4$ lines. (c-g) Optical images of the as-grown ribbons. The ribbon width reduction reflects gradual consumption of Mo from liquid Na$_2$MoO$_4$ droplet during growth. The bright dots terminating individual ribbons are circled. They indicate that the ribbons were grown from "liquid catalyst" in VLS mode.

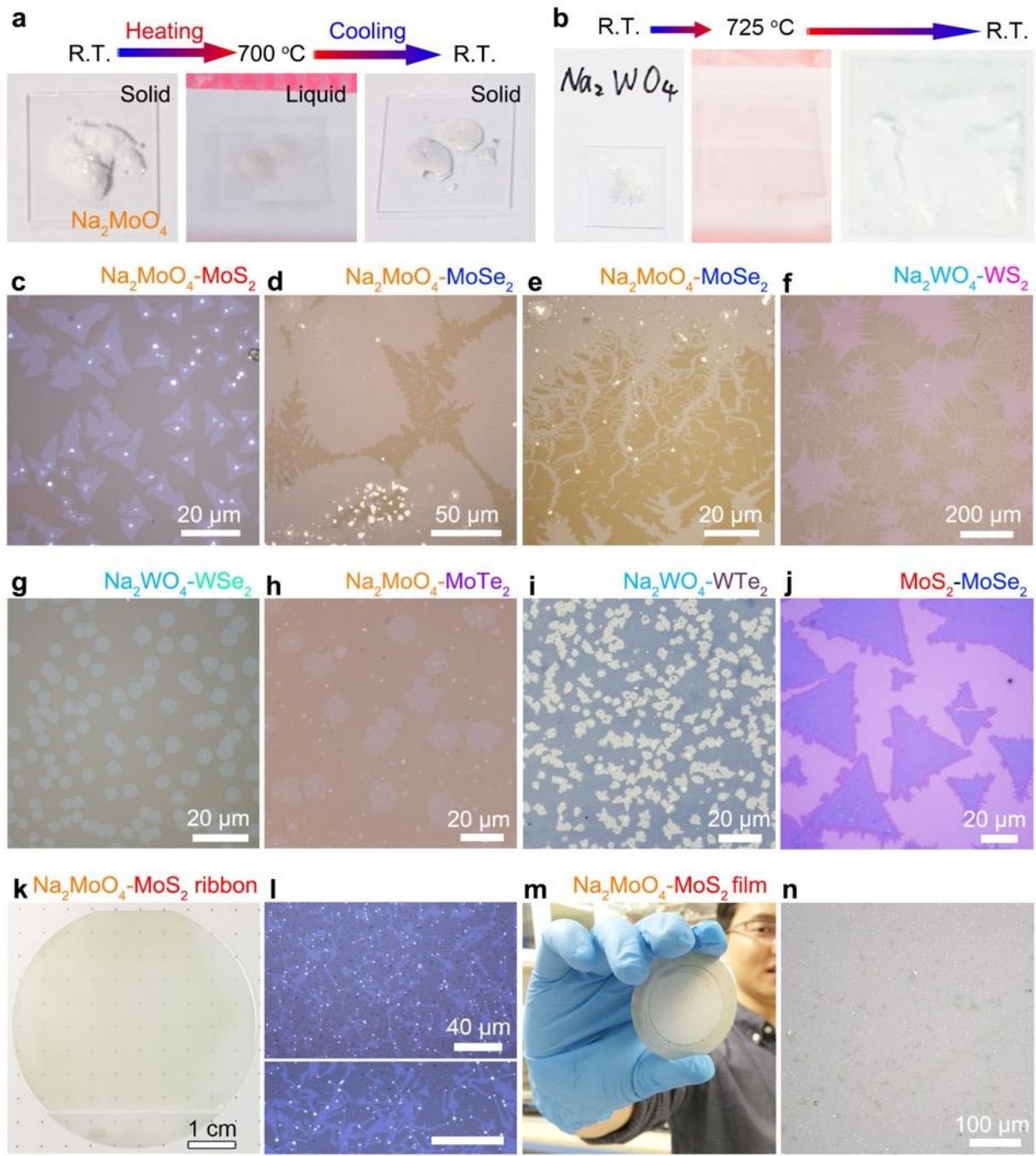

**Supplementary Figure 8 | Growth of TMDs with Na₂MoO₄ and Na₂WO₄ as precursors.** (a) Direct observation of the melting of Na₂MoO₄ when annealing at 700 °C in Ar. (b) Direct observation of the melting of Na₂WO₄ when annealed at 725 °C in Ar. (c) Optical image of monolayer MoS₂ crystals grown with Na₂MoO₄ at 700 °C in Ar. (d, e) Optical images of monolayer MoSe₂ crystals grown with Na₂MoO₄ at 725 °C in Ar. (f, g) Optical images of monolayer (f) WS₂ and (g) WSe₂ crystals grown with Na₂WO₄ at 750 °C in Ar/H₂ (5%), respectively. (h, i) Optical image of monolayer (h) MoTe₂, (i) WTe₂ crystals grown with

Na$_2$MoO$_4$ and Na$_2$WO$_4$ at 750 °C in Ar/H$_2$ (5%), respectively. (j) Optical image of MoS$_2$/MoSe$_2$ heterostructure grown with Na$_2$MoO$_4$ at 725 °C for MoSe$_2$ on monolayer MoS$_2$ flakes. (k, l) Growth of MoS$_2$ ribbons on 2-inch sapphire substrate with Na$_2$MoO$_4$ at 775 °C. (m, n) Growth of MoS$_2$ film on 2-inch sapphire substrate with Na$_2$MoO$_4$ at 725 °C.

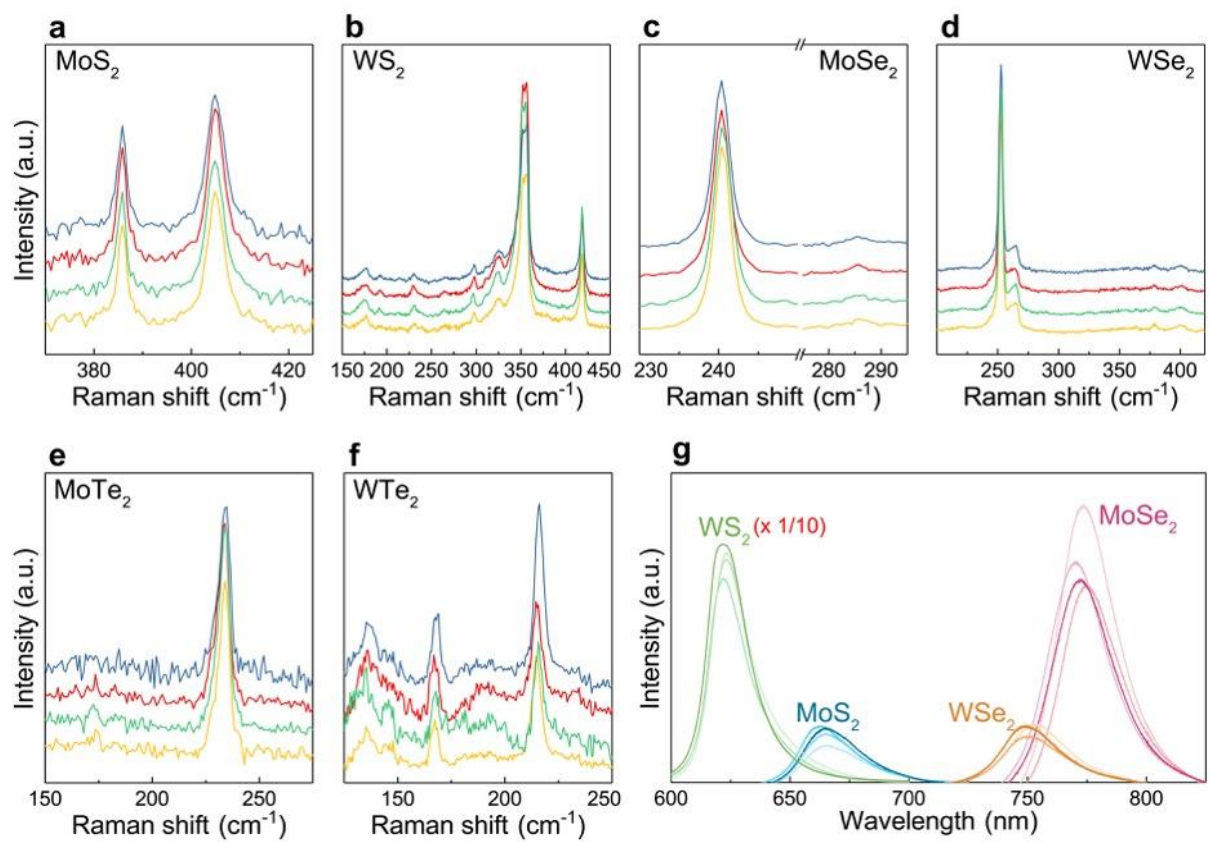

**Supplementary Figure 9 | Raman and PL spectra of as-grown TMDs with Na$_2$MoO$_4$ and Na$_2$WO$_4$ as precursors.** Raman spectra of as-grown (a) MoS$_2$, (b) WS$_2$, (c) MoSe$_2$, (d) WSe$_2$, (e) MoTe$_2$, (f) WTe$_2$. (g) PL spectra of WS$_2$, MoS$_2$ WSe$_2$ and MoSe$_2$ monolayers.

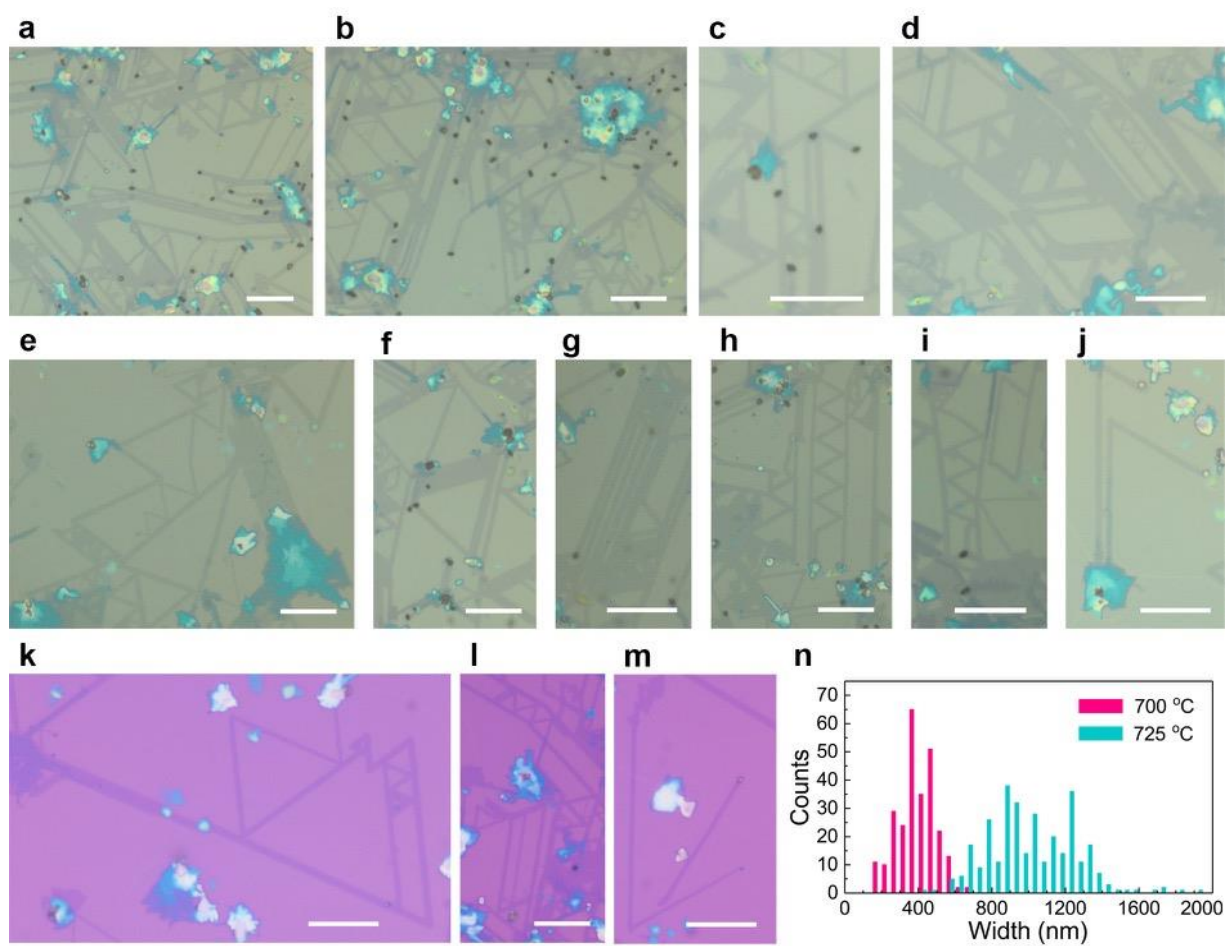

**Supplementary Figure 10 | Optical images of MoS$_2$ ribbons and islands grown on monolayer MoS$_2$.** All the samples were grown on MoS$_2$/SiO$_2$ (285 nm)/Si substrates. Figures a-j show different colors due to the cover of a PMMA layer. The scale bar is 10 μm. (n) Width distribution of as-grown MoS$_2$ ribbons. The MoS$_2$ nanoribbons grown at 700 °C have an average width of ~400 nm. The MoS$_2$ ribbons grown at 725 °C have an average width of ~1020 nm.

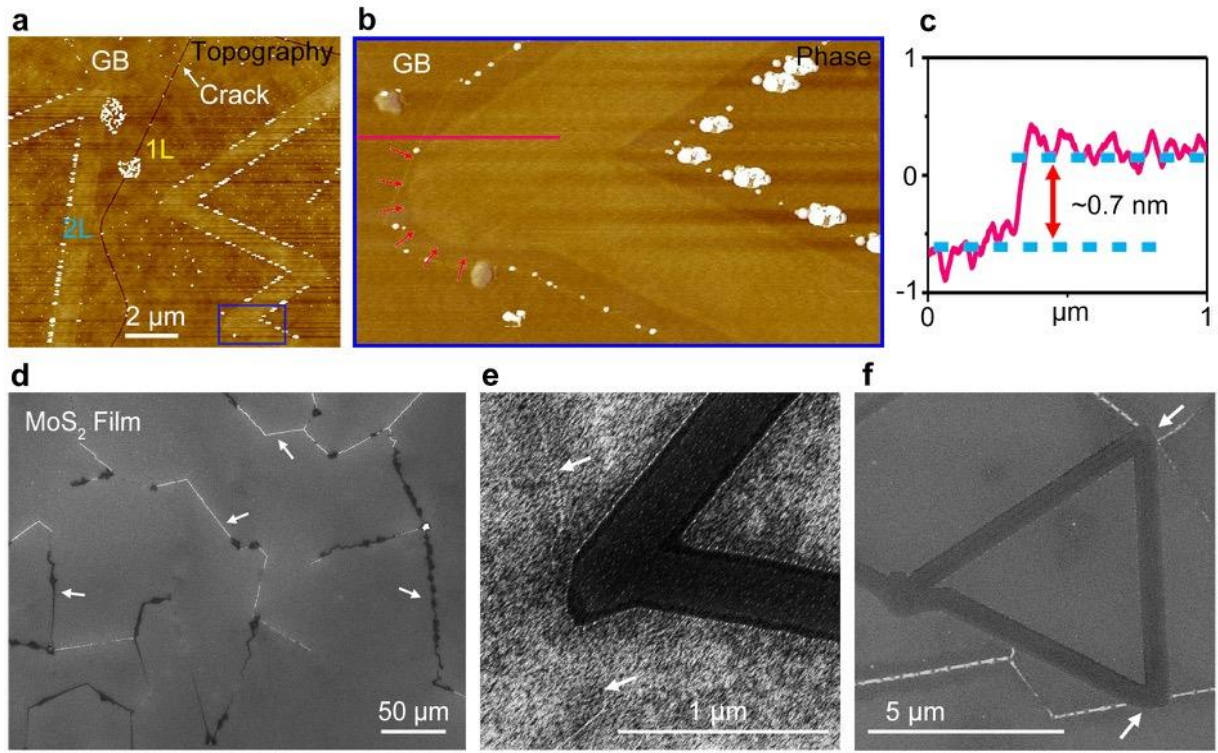

**Supplementary Figure 11 | MoS₂ ribbon kinks and the GBs of the monolayer MoS₂.** (a) AFM topography image of two MoS$_2$ ribbons grown on two adjacent MoS$_2$ grains separated by a grain boundary (GB). (b) AFM phase image showing the kink region. (c) The height profile along the red line in (b). The step height of ~0.7 nm indicates that the MoS$_2$ ribbon is monolayer. (d) SEM image revealing line features appearing at the GBs of a monolayer MoS$_2$ film. (e) SEM image showing kinks of MoS$_2$ ribbons coinciding with the GBs. (f) GBs after partially etching with H$_2$O vapor. The line features of GBs are more clearly visible.

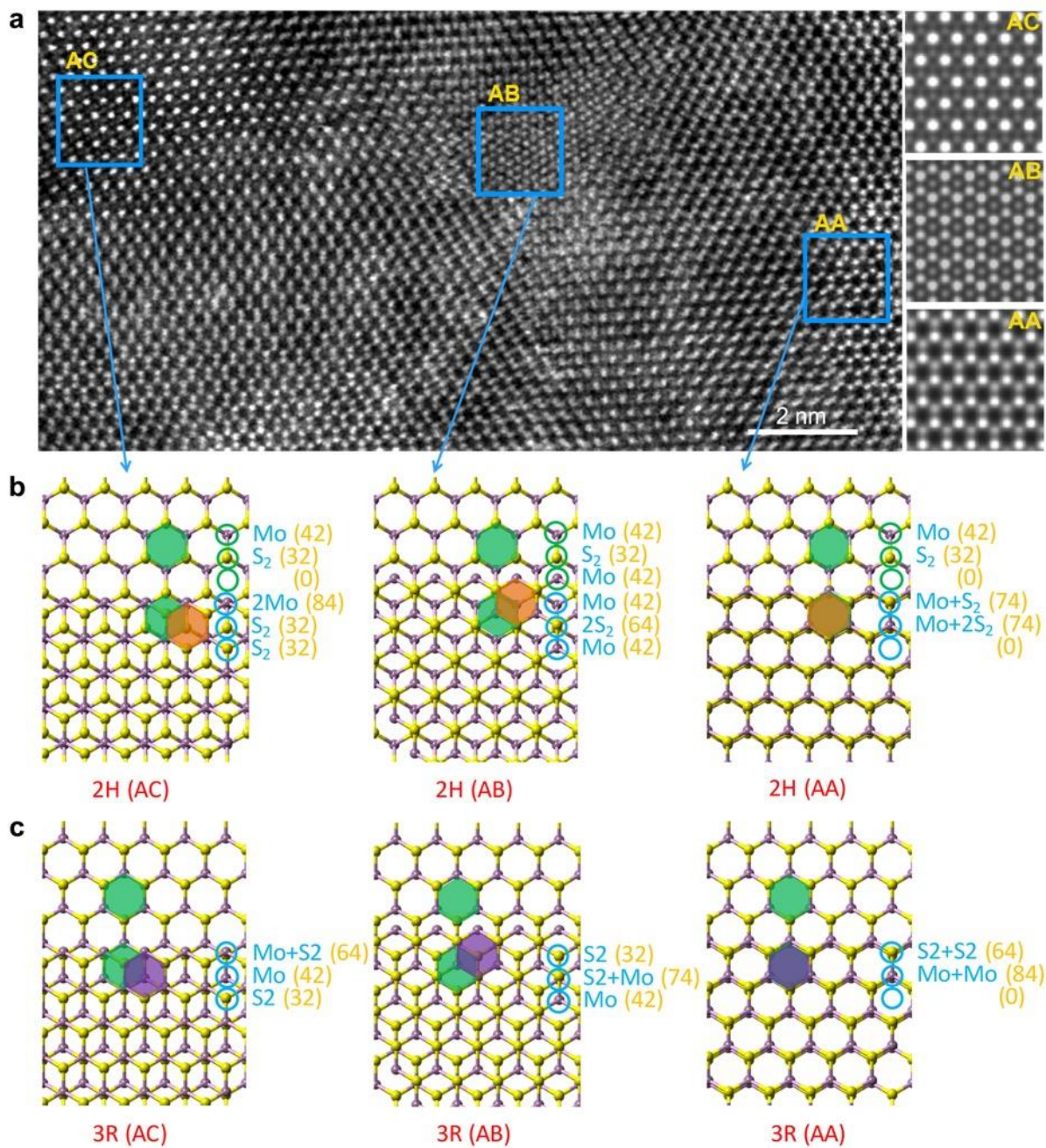

**Supplementary Figure 12 | ADF-STEM analysis of the center of a MoS$_2$ nanoribbon grown on monolayer MoS$_2$.** (a) STEM image showing spatially varying stacking sequence in the center region of a 2H-dominated MoS$_2$ nanoribbon. (b, c) Schematic illustration of AC, AB and AA stacking sequence in (b) 2H-type and (c) 3R-type MoS$_2$ nanoribbons. The number correspond to the expected ADF-STEM intensity variations. The ADF-STEM intensity variations in (a) are consistent with the variation expected in 2H-type stacking with relative displacement of the layers.

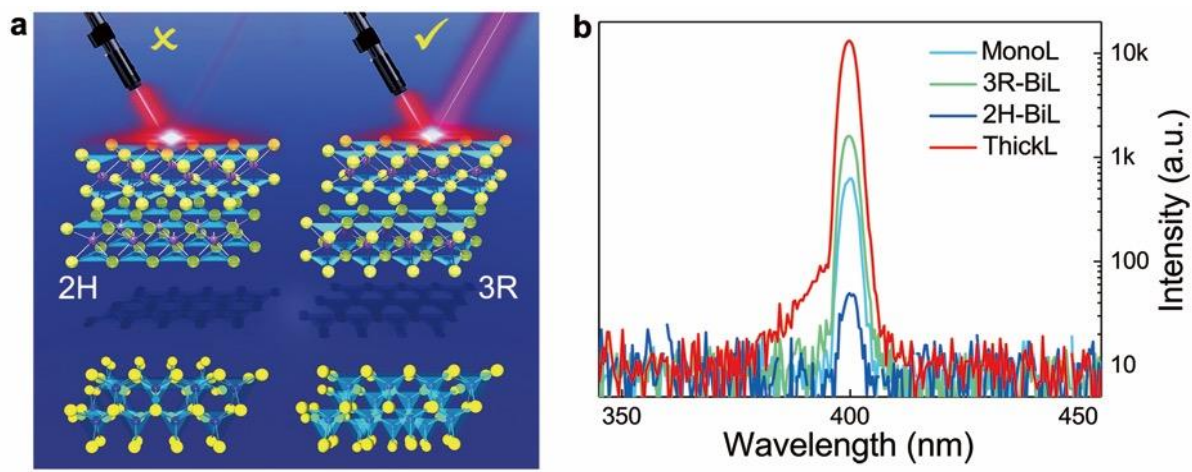

**Supplementary Figure 13 | SHG of MoS$_2$ nanoribbons grown on monolayer MoS$_2$.** (a) Schematic illustration of SHG of bilayer 2H- and 3R-MoS$_2$. (b) SHG spectra of monolayer MoS$_2$, bilayer 2H- and 3R-MoS$_2$ nanoribbons and thick MoS$_2$ islands.